\documentclass[aps,amsmath,amssymb,nofootinbib,preprintnumber,showpacs,12pt]{iopart}

\usepackage{epsfig}
\usepackage{bbm}
\usepackage{graphicx}
\usepackage{ragged2e}
\usepackage{datetime}
\usepackage{color}
\usepackage{amssymb}
\usepackage{bbold}

\newif\iffigs
\figstrue

\newcommand{\ket}[1]{\ensuremath{| #1 \rangle}}
\newcommand{\bra}[1]{\ensuremath{\langle #1 |}}
\newcommand{\nn}{\ensuremath{\nonumber}}
\newcommand{\s}[1]{\ensuremath{\sigma}^{#1}}

\begin{document}

\title{Simulating open quantum systems: from many-body interactions to stabilizer pumping}

\author{M M\"uller$^{1,2}$
, K Hammerer$^{1,2,3}$, Y L Zhou$^{1,4}$, C F Roos$^{2,5}$ and P Zoller$^{1,2}$}

\address{$^1$ Institut f\"ur Theoretische Physik, Universit\"at Innsbruck, Technikerstr. 25, 6020 Innsbruck, Austria}

\address{$^{2}$ Institut f\"ur Quantenoptik und Quanteninformation, \"Osterreichische Akademie der Wissenschaften, Technikerstr. 21A, 6020 Innsbruck, Austria}

\address{$^{3}$ Institut f\"ur Theoretische Physik, Universit\"at Hannover, Appelstr. 2, 30167 Hannover, Germany}

\address{$^{4}$ College of Science, National University of Defense Technology, Changsha, 410073, China}

\address{$^{5}$ Institut f\"ur Experimentalphysik, Universit\"at Innsbruck, Technikerstr. 25, 6020 Innsbruck, Austria}

\ead{markus.mueller@uibk.ac.at}

\begin{abstract}
In a recent experiment, Barreiro et al.~demonstrated the fundamental building blocks of an open-system quantum simulator with trapped ions [Nature 470, 486 (2011)]. Using up to five ions, single- and multi-qubit entangling gate operations were combined with optical pumping in stroboscopic sequences. This enabled the implementation of both coherent many-body dynamics as well as dissipative processes by controlling the coupling of the system to an artificial, suitably tailored environment. This engineering was illustrated by the dissipative preparation of entangled two- and four-qubit states, the simulation of coherent four-body spin interactions and the quantum non-demolition measurement of a multi-qubit stabilizer operator. In the present paper, we present the theoretical framework of this gate-based (``digital") simulation approach for open-system dynamics with trapped ions. In addition, we discuss how within this simulation approach minimal instances of spin models of interest in the context of topological quantum computing and condensed matter physics can be realized in state-of-the-art linear ion-trap quantum computing architectures. We outline concrete simulation schemes for Kitaev's toric code Hamiltonian and a recently suggested color code model. The presented simulation protocols can be adapted to scalable and two-dimensional ion-trap architectures, which are currently under development.
\end{abstract}



\maketitle


\section{Introduction}


In view of the inherent difficulties to efficiently simulate quantum physics of an interacting many-body quantum system on a classical computer due to the Hilbert space growing exponentially with the system size, Feynman proposed the idea of a quantum simulator. He suggested a controllable quantum device to efficiently study the dynamics of another quantum system of interest \cite{T_feynman-IJTPhys-21-467}. This idea was later refined and formally developed by Lloyd \cite{T_lloyd-science-273-1073} and others, who showed that many-body quantum systems can indeed be simulated efficiently, as long as they evolve according to local interactions. Since then, quantum simulation has become a very active and rapidly evolving research field on its own (see references \cite{T_buluta-science-326-108,T_monroe-physics-world-21-32} for a recent overview). Driven by remarkable experimental progress and novel theoretical ideas for various physical platforms in recent years, in particular AMO systems ranging from cold atoms \cite{T_bloch-rmp-2008,T_greiner-nature-415-39,T_trotzky-science-319-295,T_jaksch-prl-81-3108} and polar molecules \cite{T_njp-focus-polarmolecules,T_micheli-natphys-2-341} over trapped ions \cite{T_blatt-nature-453-1008} to photonic setups \cite{T_obrien-science-318-1567,T_lanyon-nat-chem-2-106,T_ma-arxiv:1008.4116} and nuclear magnetic resonance \cite{T_baugh-physics-in-canada-63-197} have been under investigation for quantum simulation. Similar promising developments have been reported for solid-state systems \cite{T_hanson-nature-453-1043} such as, e.g.~, arrays of coupled superconductors \cite{T_clarke-nature-453-1031,T_pritchett-arxiv-1008.0701}, quantum dots \cite{T_hanson-rmp-79-1217} and nitrogen-vacancy centers in diamond \cite{T_wrachtrup-jpcm-18-S807}. 


For \textit{closed} many-body quantum systems, which are well-isolated from their environment, powerful techniques have been developed to control the internal, coherent dynamics. The ability to engineer and tune the underlying single-particle and interaction Hamiltonian terms has enabled the simulation of different classes of quantum many-body models over wide ranges of parameters. Ultimately, though, every quantum system is unevitably coupled also to its surrounding environment. Recently, quantum control of \textit{open} many-body systems, which amounts to engineering both the Hamiltonian time evolution of the many-body system itself as well as its coupling to the environment \cite{T_lloyd-science-273-1073,T_lloyd-pra-65-010101}, has become a major research focus. Whereas typically the system-environment coupling leads to detrimental effects on many-body or multi-qubit open systems \cite{T_myatt-nature-403-269,T_viola-science-293-2059,T_deleglise-nature-455-510,T_barreiro-natphys-aop}, the ability to control and tailor the associated dissipative processes has been identified as a useful resource: it allows one to dissipatively prepare entangled quantum states and correlated quantum phases from arbitrary initial states \cite{T_krauter-arxiv-1006.4344,T_muschik-arxiv-1007.2209,T_cho-prl-106-020504,T_diehl-natphys-4-878,T_kraus-pra-78-042307,T_herdman-prl-104-230501}, and can also be exploited for dissipative quantum computing \cite{T_verstraete-nphys-5-633} and quantum memories \cite{T_pastawski-arxiv:1010.2901}.

\begin{figure}[ht]
\begin{center}
\includegraphics[width=0.95\columnwidth]{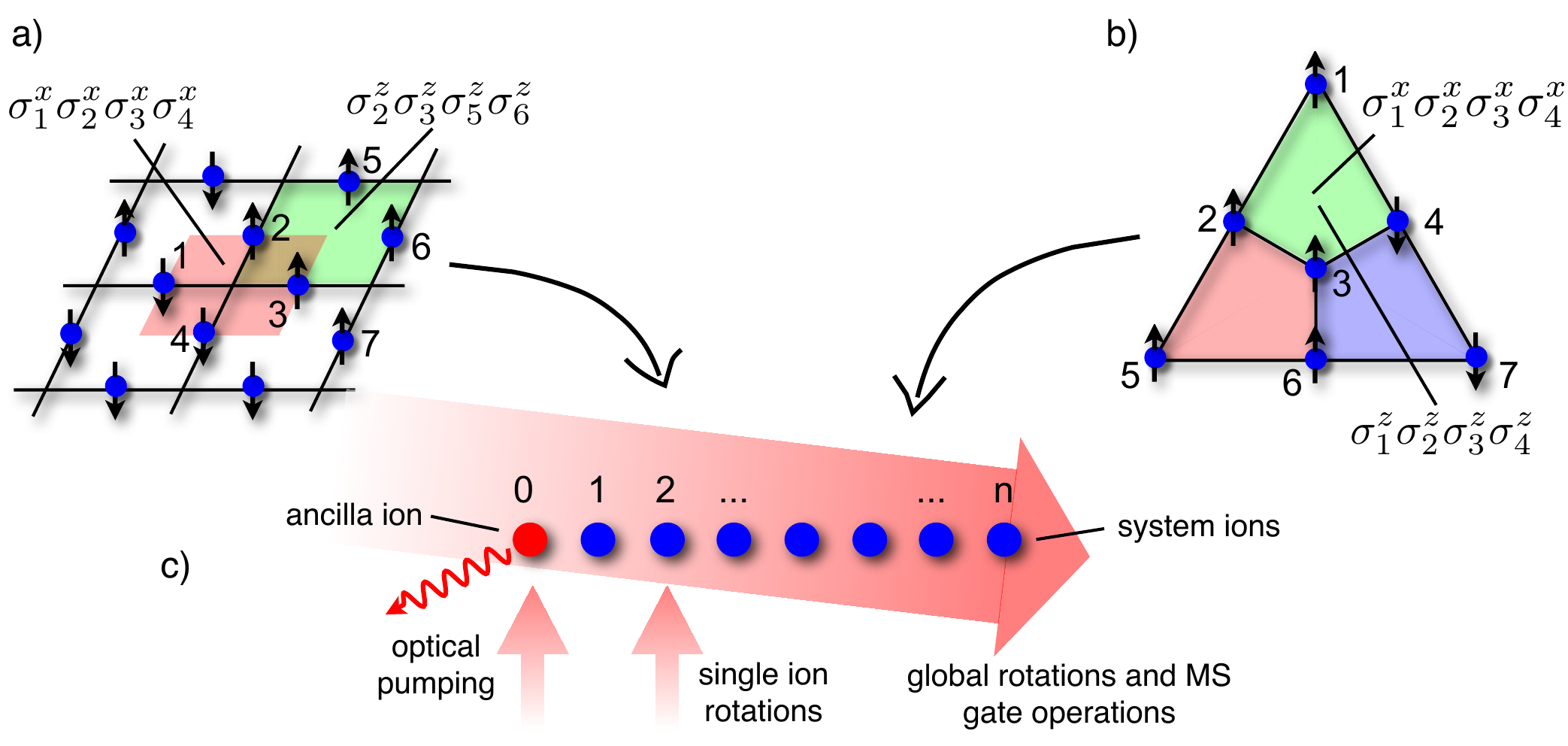}
\end{center}
\caption{Lattice spin models of interest for the gate-based ("digital") quantum simulation with trapped ions.  (a) In Kitaev's toric code \cite{T_Kitaev-annalsphys-303-2} spins located around vertices of a two-dimensional square lattice interact via four-body interactions $\sim \s{x}_1\s{x}_2\s{x}_3\s{x}_4$, whereas spins around plaquettes experience $z$-type interactions, as e.g.~$\sim \s{z}_2\s{z}_3\s{z}_5\s{z}_6$. (b) Small instance of a \textit{color code} spin system, as proposed in \cite{T_bombin-prl-97-180501}. Here, spins are located on the sites of a three-colorable lattice interact via four-body plaquette interactions such as $\sim \s{x}_1\s{x}_2\s{x}_3\s{x}_4$ and $\sim \s{z}_1\s{z}_2\s{z}_3\s{z}_4$. (c) Mesoscopic instances of spin models can be mapped onto linear chains of trapped ions, where the spin degree of freedom is encoded in (meta)stable electronic states. Coherent and dissipative time evolution can be simulated by sequences of highly parallel multi-ion M\o lmer-S\o rensen (MS) gates applied to all (or subsets of) ions, in combination with single-qubit rotations on individual ions and optical pumping of an ancilla ion.}
\label{fig:latticemodels}
\end{figure}

Recently, the elementary building blocks of such an open-system quantum simulator have been shown in an experiment with up to five ions \cite{T_barreiro-nature-2011}. In their work, Barreiro \textit{et al.} demonstrated the ability to engineer coherent and dissipative multi-qubit quantum operations by the dissipative preparation of Bell states and multi-qubit stabilizer states, the simulation of coherent four-body spin interactions and a quantum non-demolition measurement of four-qubit stabilizer operators. Since the theoretical concepts and details of this work are of general interest to the ion trap community in the context of quantum simulation of spin systems, we provide in the present paper the theoretical framework of the simulation scheme. The present work is motivated by the developments of ideas in the context of topological spin models in the context of quantum computing and condensed matter and the question to what extent these ideas can be realized in existing experimental setups, in particular with linear ion-trap architectures. 
We focus on the following questions: What are interesting simulation possibilities in state-of-the-art ion trap quantum computing setups with moderately large chains of a few, possibly up to a few tens of ions? And how can the currently available experimental resources be exploited in an optimal and \textit{experimentally efficient} way that allows one to access the physics of minimal instances of complex spin models (as schematically shown in figure \ref{fig:latticemodels}) with today's technology?
%


Below, we present a toolbox for the simulation of general Markovian open-system dynamics of mesoscopic spin systems. Our set of tools includes the fundamental building blocks for  the simulation of coherent $n$-body spin interactions, dissipative $n$-qubit reservoir engineering, and the ability for quantum-non-demolition (QND) measurements of $n$-particle observables. The simulation scheme strongly makes use of the well-developed set of tools for the purpose of quantum state preparation, manipulation and readout of trapped ions \cite{T_haffner-physrep-469-155,T_blatt-nature-453-1008,T_home-science-325-1227}. In particular, we show how high-fidelity multi-ion MS entangling gate operations \cite{T_molmer-prl-82-1835}, as first suggested M\o lmer and S\o rensen, and recently shown for up to 14 ions in the laboratory \cite{T_monz-superdecoherence}, conveniently bundle the effect of sequences two-qubit operations. This allows one to reduce the experimental simulation complexity significantly and to realize, e.g., coherent $n$-body interactions in a minimal number of steps.
In our simulation architecture, we use optical pumping on individual ions - in combination with coherent gate operations - to tailor the coupling of the spin system to its environment and thereby engineer dissipative $n$-body quantum processes. 


Our ``digital" simulation scheme is based on the stroboscopic application of  sequences of coherent gate operations in combination with dissipative time steps to realize open-system dynamics. It complements existing proposals of quantum simulation with ground state ions \cite{T_porras-prl-92-207901,T_jane-quant-inf-comp-3-15,T_cirac-prl-74-4091} or ions excited to Rydberg states \cite{T_mueller-njp-9-093009}. In these ``analog" quantum simulators, the common principle is to use externally controllable fields to engineer effective ``always-on" Hamiltonians, which microscopically realize the model of interest directly. Recently, remarkable first experiments have demonstrated the simulation of (relativistic) single-particle dynamics in an external potential \cite{T_leibfried-prl-89-247901,T_gerritsma-nature-463-68,T_gerritsma-prl-106-060503} and experimental studies of the physics of few interacting Ising spins \cite{T_friedenauer-nphys-4-757} under frustration \cite{T_kim-nature-465-590}.


We point out that the presented ``digital" simulation scheme is suited for the simulation of mesoscopic spin systems corresponding to chains of up to a few tens of ions, which with state-of-the art ion trap technology can be controlled accurately. However, similar protocols can be realized in scalable and two-dimensional ion-trap architectures, to whose development currently a lot of effort is devoted \cite{T_blakestad-prl-102-153002,T_home-science-325-1227,T_hensinger-apl-88-034101,T_schmied-prl-102-233002,T_clark-j-appl-phys-105-013114}, and also on other physical simulation platforms. In fact, in previous work a ``digital" quantum simulation architecture for open-system dynamics of many-body spin models has been developed for neutral Rydberg atoms in optical lattices \cite{T_weimer-nphys-6-382}.


In section \ref{sec:idea_and_results} we introduce the general idea of our simulation architecture and give a concise summary of the main results. The details of the simulation of coherent and dissipative many-body interactions are provided in sections \ref{sec:coh_simulation} and \ref{sec:diss_simulation}. In section \ref{sec:imperfections} we briefly discuss the effect of imperfections in the simulation scheme. We illustrate our simulation scheme in section \ref{sec:applications} for two examples of interest in the context of topological quantum computing, namely small-scale implementations of Kitaev's  toric code \cite{T_Kitaev-annalsphys-303-2} and a minimal instance of a color-code model \cite{T_bombin-prl-97-180501}. We conclude with an outlook. 

\section{Simulation of open many-body quantum systems}
\label{sec:idea_and_results}

\subsection{Open-system dynamics}
In the following we are interested in open-system dynamics of many-body quantum systems. The dynamics of an open quantum system which is coupled to an environment can be described by a completely positive Kraus map \cite{T_nielsen-book}
\begin{equation}
\label{eq:Kraus_map}
\rho \mapsto \sum_k E_k \rho E_k^\dagger,
\end{equation}
where $\rho$ denotes the reduced density operator of the system, $\{E_k\}$ is a set of operation elements satisfying $\sum_k E_k^\dagger E_k = 1$, and we assume an initially uncorrelated system and environment. For the case of a closed system, decoupled from the environment, the map (\ref{eq:Kraus_map}) reduces to $\rho \mapsto U \rho U^\dagger$ with $U$ the unitary time evolution operator of the system. 

In the literature on quantum control of open quantum systems, the required set of operations to realize different classes of quantum operations (\ref{eq:Kraus_map}) as well as efficiency and universality aspects have been discussed \cite{T_lloyd-pra-65-010101,T_bacon-pra-64-062302}. In reference \cite{T_barreiro-nature-2011}, several specific examples of Kraus maps, whose dissipative dynamics can be used for dissipative quantum state preparation, e.g.~Êfor pumping into entangled states, have been discussed and implemented experimentally.

The Markovian limit of the general quantum operation (\ref{eq:Kraus_map}) for the \emph{coherent} and \emph{dissipative} dynamics of a many-particle spin system is given by a many-body master equation
\begin{equation}
\label{eq:master_equation}
\frac{\mathrm{d}}{\mathrm{d}t} \rho
 = (\mathcal{L}^{\mathrm{coh}}+\mathcal{L}^{\mathrm{diss}}) \rho
 = \mathcal{L} \rho
\end{equation}
for the density operator $\rho(t)$ of the many-body system. The \emph{coherent} part of the dynamics is described by
\begin{equation}
\label{eq:unitary_propagator}
\mathcal{L}^{\mathrm{coh}} \rho = \sum_\alpha \mathcal{L}_\alpha ^{\mathrm{coh}} \rho = -i \sum_\alpha [H_\alpha, \rho ].
\end{equation}
It is generated from a Hamiltonian $H = \sum_\alpha H_\alpha$ which is a sum of terms $H_\alpha$, which can in general involve higher order $n$-body interactions, which act on a quasi-local subset of particles \footnote{Throughout this article we set $\hbar=1$.}. \emph{Dissipative} time evolution is described by a Liouvillian part of the master equation
\begin{equation}
\label{eq:Lindblad}
\mathcal{L}^{\mathrm{diss}} \rho = \sum_\alpha \mathcal{L}_\alpha ^{\mathrm{diss}} \rho \nonumber  = \sum_\alpha \frac{\gamma_\alpha}{2} \left( 2c_\alpha \rho c_\alpha^\dagger - c_\alpha^\dagger c_\alpha \rho
-  \rho c_\alpha^\dagger c_\alpha  \right).
\end{equation}
The individual terms $\mathcal{L}_\alpha ^{\mathrm{diss}} \rho$ are of Lindblad form \cite{T_wiseman-book}, and are determined by quantum jump operators $c_\alpha$, which either act on single or on subsets of particles, and by respective rates $\gamma_\alpha$ at which these jump processes occur. Engineering open-system dynamics thus amounts to designing and engineering couplings of the quantum system to its environment, such that the resulting many-particle dynamics is then governed by discrete Kraus maps or master equations with quasi-local Hamiltonian and dissipative terms.

\subsection{Many-body quantum systems: Kitaev's toric code as a representative example}
In the following, we will in consider the simulation of many-body lattice spin models, which are of interest in the context of topological quantum computing and memories. As a paradigmatic example of this class of spin models we discuss in some detail Kitaev's toric code Hamiltonian, which is sketched in figure \ref{fig:latticemodels}(a). This model exemplifies in a transparent way the challenges that one encounters also in the quantum simulation of related models, such as e.g.~in a recently suggested color code model (see figure \ref{fig:latticemodels}(b)), which we discuss in more detail in section \ref{sec:colorcode}.

In Kitaev's toric code model, as sketched in figure \ref{fig:latticemodels}(a), spins are located on the edges of a two-dimensional square lattice. The Hamiltonian is given by $H = - E ( \sum_\mathrm{s} A_\mathrm{s} + \sum_\mathrm{p} B_\mathrm{p})$, which is a sum of stabilizer operators 
\begin{equation}
\label{eq:Kitaev_operators}
A_s = \prod_{i \in s} \s{x}_i \qquad \mathrm{and} \qquad B_p= \prod_{i \in p} \s{z}_i, 
\end{equation}
which describe four-body interactions of spins, which are located around the vertices (stars) $s$ and plaquettes $p$ of the lattice, respectively. All four-body stabilizers have eigenvalues $\pm 1$ and mutually commute. The ground state(s) is/are thus given by the simultaneous eigenstate(s) of all stabilizers with eigenvalues $+1$ (assuming $E_\mathrm{s}, E_\mathrm{p}>0$). The degeneracy of the ground state depends on the boundary conditions and topology of the setup. Excited states in this model correspond to violations of these stabilizer constraints, i.e., $-1$ eigenstates with respect to either the $A_s$ or $B_p$ stabilizers. They can be associated with localized quasiparticles residing on the corresponding vertices and plaquettes of the lattice (as illustrated in figure \ref{fig:2KitaevPlaquettes}(b)). They exhibit anyonic statistics under braiding, i.e.~when trajectories of different types of quasiparticles are winded around one another. 

Preparation of the system in the ground state manifold, starting from an arbitrary initial (excited) state, can be achieved by a dissipative dynamics which is governed by a many-body master equation (\ref{eq:master_equation}) with quantum jump operators 
\begin{equation}
c_\alpha = \frac{1}{2} \sigma_i^z (1- \sigma_1^x \sigma_2^x \sigma_3^x \sigma_4^x) \qquad \mathrm{and} \qquad c_\beta = \frac{1}{2} \sigma_i^x (1- \sigma_1^z \sigma_2^z \sigma_3^z \sigma_4^z) 
\end{equation}
\begin{figure}[h!]
\begin{center}
\vspace{3mm}
\includegraphics[width=0.75\columnwidth]{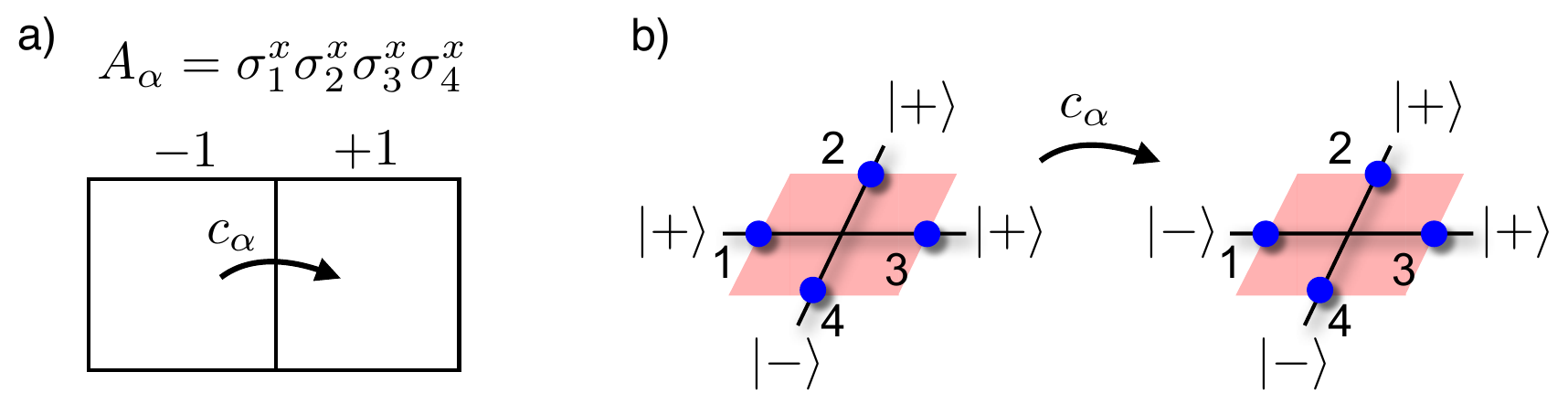}
\end{center}
\caption{Illustration of the dissipative dynamics of stabilizer pumping of four spins: Lindblad dynamics according to a four-body quantum jump operator $c_\alpha =  \frac{1}{2} \sigma_1^z (1- \sigma_1^x \sigma_2^x \sigma_3^x \sigma_4^x)$ induces pumping into the eigenspace of eigenvalue +1 of the four-body stabilizer operator $A_\alpha = \sigma_1^x \sigma_2^x \sigma_3^x \sigma_4^x$. All +1 eigenstates are left invariant, whereas eigenstates corresponding to an eigenvalue -1 of $A_\alpha$ are incoherently converted into +1 eigenstates, e.g.~$c_\alpha |+++-\rangle = |-++-\rangle$, by a flip of one of the four spins.}
\label{fig:plaquette_cooling}
\end{figure}
These collective operators act on four spins located around a vertex (site) of the lattice, as depicted in figure \ref{fig:latticemodels}(a). The index $i$ denotes one arbitrary spin of the four involved spins. A four-body jump operator $c_\alpha$ induces dissipative dynamics, which pumps the four spins from the +1 into the -1 eigenspace of  $A_\alpha = \sigma_1^x \sigma_2^x \sigma_3^x \sigma_4^x$ (see figure \ref{fig:plaquette_cooling}). The projector part $\frac{1}{2} (1- \sigma_1^x \sigma_2^x \sigma_3^x \sigma_4^x)$ applied to any $+1$ eigenstate of $A_\alpha$ vanishes; as a consequence all +1 eigenstates are ``dark states" and remain unaffected. In contrast, the spin flip $\sigma_i^z$ applied to one of the four spins (e.g.~$i=1$) can incoherently convert -1 into +1 eigenstates, e.g.,~$c_\alpha |+++-\rangle = \sigma_1^z|+++-\rangle = |-++-\rangle$. Here, $\ket{\pm}$ are the eigenstates of $\sigma^x$: $\sigma^x \ket{\pm} = \pm \ket{\pm}$. 

The above example illustrates that the difficulty to be overcome in simulating the coherent Hamiltonian dynamics lies in finding a way to realize the four-body Hamiltonian interaction terms. The realization of the dissipative ``cooling" dynamics into the ground state(s) by means of the described collective dissipative processes requires the engineering of a coupling of the spin system to an artificial, tailored environment. An \textit{analog} simulation of these coherent and dissipative higher-order $n$-body interactions, i.e.,~by a direct engineering using ``always-on" external fields, is demanding because these higher-order effective interactions must be constructed from underlying one- and two-body interactions. In this scenarion, typically, the interaction strengths and dissipative rates of the $n$-body processes, which typically arise in a perturbative limit, are much smaller than dominant one- and two-body interactions. 

Therefore, we aim to realize the coherent and dissipative dynamics according to (\ref{eq:Kraus_map}) or (\ref{eq:master_equation}) in a \textit{digital} simulation, i.e.~by stroboscopic sequences of gates and dissipative operations. Here, higher-order $n$-body interactions can be obtained non-perturbatively as leading-order terms from the application of one-, two- or $n$-body quantum gates. The corresponding interaction strengths are virtually independent of the order $n$ of the interaction terms and ultimately only limited by the gate durations in the underlying quantum circuits. 

In the case of continuous time dynamics, we apply these operations over small time steps $\tau$, such that the master equation (\ref{eq:master_equation}) emerges as an effective, coarse-grained description of the time evolution. For small time steps, the time evolution can be implemented through a Trotter expansion of the propagator corresponding to Eq.~(\ref{eq:master_equation})
\begin{equation}
\label{eq:Trotter_expansion}
e^{\mathcal{L}\tau} \rho \simeq \prod_{\alpha} e^{\mathcal{L}_\alpha^{\mathrm{coh}}\tau} \prod_{\beta} e^{\mathcal{L}_\beta^{\mathrm{diss}}\tau} \rho.
\end{equation}
Errors from possible non-commutativity of the quasi-local terms in $\mathcal{L}$ are bounded \cite{T_nielsen-book} and can be reduced by resorting to shorter time steps $\tau$ or higher-order Trotter expansions \cite{T_suzuki-pla-165-387}. 
On the other hand, as we will discuss below, it is also possible to engineer sequences of discrete Kraus maps (\ref{eq:Kraus_map}), which can for instance be employed for dissipative quantum state preparation in a minimal number of steps. 

\subsection{Experimental tools for digital quantum simulation with trapped ions}

\begin{figure}[ht]
\begin{center}
\includegraphics[width=0.95\columnwidth]{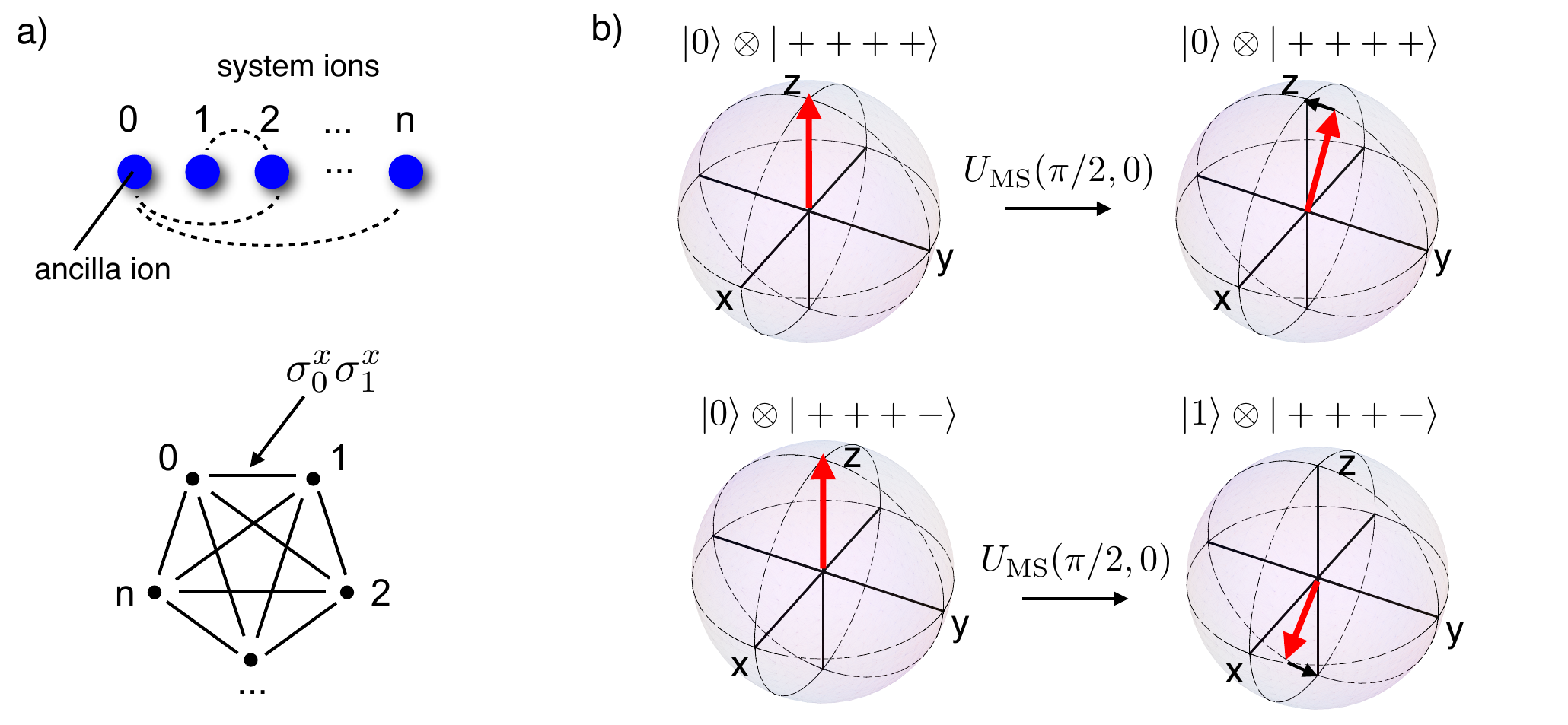}
\end{center}
\caption{(a) Graph representation of the two-body spin interaction Hamiltonian, which underlies the multi-ion MS gate (\ref{eq:MS_gate}). All pairs $\{i,j\}$ of ions involved in the gate interact with equal strength (represented as links). (b) A $(4+1)$ ion entangling MS gate applied to $4$ system ions and an ancilla ion (index zero) can be used to coherently map the information about whether the $4$ system ions are in a +1 (-1) eigenstate of the $4$-body interaction term $\sim \s{x}_1\s{x}_2 \s{x}_3 \s{x}_4$, onto the logical states $\ket{0}$ and $\ket{1}$ of the ancilla ion. In the Bloch sphere representation this mapping can be understood as a rotation of the ancilla qubit initially prepared in $\ket{0}$ around the $x$-axis. The rotation angle depends on the state of the system ions, and is chosen such that for any +1 eigenstate, as e.g.~$\ket{++++}$, the ancilla qubit ends in $\ket{0}$ after the MS gate, whereas for - 1 eigenstates such as e.g.~ $\ket{+++-}$ it is transferred to $\ket{1}$. This mapping mechanism not only works for $4$-body interactions, but can be used for general $n$. For $n$ odd, the ancilla qubit is transferred to $\sigma^y$ eigenstates.}
\label{fig:bloch_sphere}
\end{figure}

Motivated by the present availability of well-developed set of coherent and dissipative tools \cite{T_barreiro-nature-2011} in state-of-the-art linear ion-trap architectures \cite{T_haffner-physrep-469-155}, we consider a setup in which the spins of a (possibly two- or three-dimensional) lattice model with a mesoscopic number of particles are mapped onto a linear chain of ions, where the spin degrees of freedom are encoded in two (meta-)stable internal states of the ions. Although our approach can be realized with any universal set of gate operations, we focus on a realization, which benefits from highly parallel multi-ion MS gates as the principal building block for the implementation of unitary and dissipative simulation time steps in eq.~(\ref{eq:Trotter_expansion}).
The MS gate operation \cite{T_molmer-prl-82-1835} is based on pairwise two-ion interaction terms (as illustrated in figure \ref{fig:bloch_sphere}), and can be parametrized by two angles $\theta$ and $\phi$,
\begin{equation}
\label{eq:MS_gate}
U_{\mathrm{MS}} (\theta, \phi) = \exp \left( -i \frac{\theta}{4}\left( \cos \phi S_x +\sin \phi S_y \right)^2 \right).
\end{equation}
The sum in the collective spin operators $S_{x,y} = \sum_{i=0}^{n} \s{x,y}_i$ with $\s{x,y}_i$ the usual Pauli matrices, is understood to be performed over all ions involved in the gate. This multi-ion entangling gate operation is complemented by (non-entangling) single- and multi-qubit rotations, whose physical implementation is discussed, e.g.,~in \cite{T_barreiro-nature-2011}. In addition to this universal set of coherent gate operations, the use of optical pumping on individual ions (as demonstrated e.g.~\cite{T_schindler-qec}) constitutes the dissipative ingredient for the engineering of dissipative many-body spin dynamics. 

\begin{figure}[h!]
\begin{center}
\includegraphics[width=0.7\columnwidth]{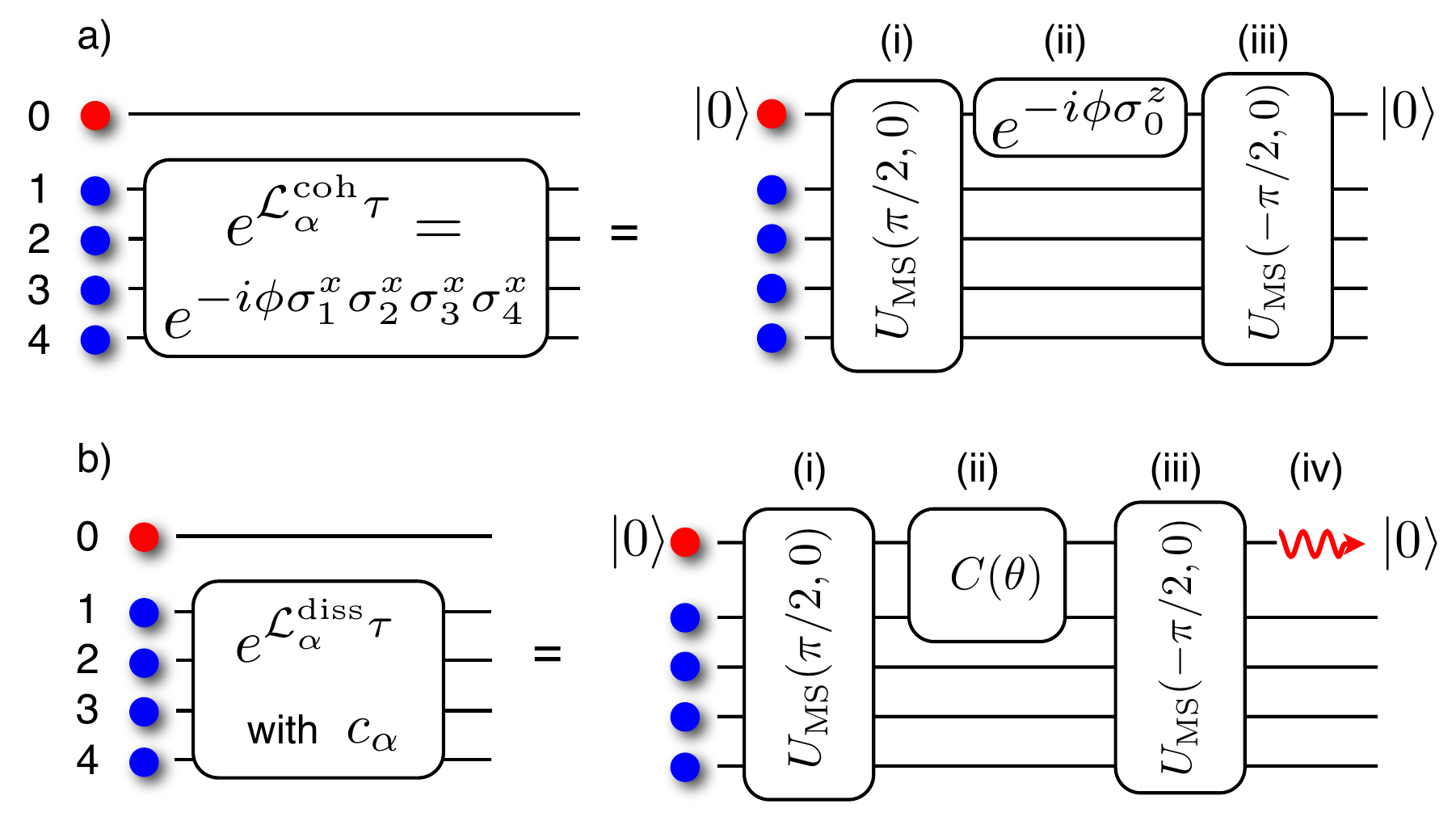}
\end{center}
\caption{Generic gate decompositions for the simulation of coherent and dissipative dynamics via the Trotter expansion (\ref{eq:Trotter_expansion}). (a) Coherent evolution according to a four-body interaction Hamiltonian $H_\alpha = E_\alpha \s{x}_1\s{x}_2\s{x}_3\s{x}_4$ for a time step $\tau$ is efficiently achieved in three steps: (i) First, an entangling MS gate $U_\mathrm{MS}(\pi/2,0)$ applied to the four system ions and the ancilla ion coherently maps the information on whether the system ions are in a +1 or -1 eigenstate of $H_\alpha$ onto the logical states $\ket{0}$ and $\ket{1}$ of the ancilla qubit (cf.~figure \ref{fig:bloch_sphere}). (ii) Second, a single-qubit gate $\exp(i\phi \s{z}_0)$ applied to the ancilla ion effectively imprints a phase $-\phi$ ($\phi$) on all +1 (-1) eigenstates of $H_\alpha$. (iii) Finally, the initial mapping is reversed by an inverse MS gate, which disentangles the ancilla from the system ions, which have evolved according to $\exp( i \phi \s{x}_1\s{x}_2\s{x}_3\s{x}_4)$. (b) Dissipative evolution, i.e.~``cooling" of the system ions into the +1 eigenspace of $H_\alpha$ mediated by Lindblad dynamics with four-body quantum jump operators $c_\alpha = \frac 12 \s{z}_1 (1 - \s{x}_1\s{x}_2\s{x}_3\s{x}_4)$: Here, a two-qubit gate $C(\theta) = \ket{0}\bra{0}_0 \otimes 1 + \ket{1}\bra{1}_0 \otimes \exp[i\theta \s{y}_i]$ is applied between the mappings (i) and (iii). In step (ii), all +1 eigenstates of $H_\alpha$ are left invariant, whereas a spin flip $\s{y}_i$ with a $\theta$-dependent amplitude is applied to one of the four system ions can convert -1 into +1 eigenstates. The angle $\theta$ controls the conversion probability and allows one to tune from small probabilities ($\theta \ll 1$, master equation limit) to unit pumping probability ($\theta = \pi/2$). After the three steps, generally, the ancilla qubit is entangled with the system ions. (iv) Finally, the ancilla qubit is incoherently reset to its initial state $\ket{0}$ by optical pumping. This dissipative step enables to carry away entropy and ``cool" the system qubits.}
\label{fig:gatesequence}
\end{figure}

We have summarized the the basic idea of the simulation of coherent and dissipative dynamics corresponding to Kitaev's code model in figure \ref{fig:gatesequence}, to be explained in more detail in the following sections.

We will show in more detail in section \ref{sec:coh_simulation} that the unitary propagators $e^{\mathcal{L}_\alpha^{\mathrm{coh}}\tau}\rho$ corresponding to $n$-body interaction Hamiltonians $H_\alpha$ (such as, e.g., the four-body term in Eq.~(\ref{eq:Kitaev_operators})) can be implemented efficiently in an experiment (i.e.~by a minimal number of gates) by combining standard single-qubit gates with $(n+1)$-ion MS gates, which are applied to the $n$ system ions and an additional ion, which encodes an ancilla qubit. Dissipative dynamics according to propagators $e^{\mathcal{L}_\alpha^{\mathrm{diss}}\tau}\rho$ with many-body jump-operators $c_\alpha$ can be achieved by combining the coherent gate operations with a dissipative step in the form of optical pumping of the ancilla ion.

%

\section{Simulation of coherent $n$-body interactions}
\label{sec:coh_simulation}

In this section we describe in more detail the stroboscopic simulation of coherent dynamics $e^{\mathcal{L}_\alpha^{\mathrm{coh}} \tau}$ according to $n$-body spin interaction Hamiltonians. 

\subsection{The M\o lmer-S\o rensen multi-ion entangling gate}
The main resource in our simulation scheme are multi-ion MS gate operations (\ref{eq:MS_gate}), which rely on the application of a bichromatic laser field to the ions \cite{T_molmer-prl-82-1835}. The two frequency components are chosen close to the qubit transition, fine-tuned such that an effective second-order coupling between pairs of ions is generated by off-resonantly coupling to the blue and red motional side-bands of the common vibrational center-of-mass mode of the ion string. Within the Lamb-Dicke regime, where the ions are spatially confined to a region much smaller than the wavelength of the qubit transition, the MS gate operation is particularly robust and works in principle without the necessity of cooling to the motional ground state \cite{T_roos-njp-10-013002}. The gate has been successfully demonstrated \cite{T_sackett-nature-404-256} with remarkably high fidelities (99.3 \% for a pair of ions \cite{T_benhelm-nphys-4-463}) and recently for strings of up to 14 ions \cite{T_monz-superdecoherence}. A detailed discussion of the MS gate, in particular regarding the experimental implementation and optimization, can be found in \cite{T_roos-njp-10-013002}. The properties of the MS gate (\ref{eq:MS_gate}) applied to $n+1$ ions, which we will repeatedly use in the following, are:
\begin{itemize}
\item
The phase $\theta$ is the main control parameter of the gate; for $\theta = \pi/2$, the gate is maximally entangling, i.e., the computational basis states are mapped to states, which are up to local rotations equivalent to GHZ states \cite{T_molmer-prl-82-1835}. Shifting the optical phase of the bichromatic driving field allows one to switch between a $\sigma^x$-type ($\phi=0$) and a $\sigma^y$-type ($\phi=\pi/2$) MS gate.
\item
Periodicity: $U_\mathrm{MS}(\theta + 2 \pi \,m,\phi) = U_\mathrm{MS}(\theta,\phi)$ for $m \in \mathbb{Z}$.
\item
``Backward" MS gates (i.e.~for negative values of $\theta$) can be realized by ``forward" gates, since
\begin{equation}
 U_\mathrm{MS}(-\theta, \phi) \equiv \left\{ \begin{array}{cc}
 U_\mathrm{MS}(\pi-\theta, \phi) &  \textrm{ for } n \textrm{ even,} \\
 U_\mathrm{MS}(\pi-\theta, \phi) \left( \prod_{j=0}^n \tilde{\sigma}_j \right) & \textrm{ for } n \textrm{ odd.}\end{array} \right.
\end{equation}
with $\tilde{\sigma}_j = \cos \phi \s{x}_j +\sin \phi \s{y}_j$. In particular, fully entangling MS gates are (up to local rotations for an odd number of ions) equivalent to their inverse operations. Using only ``forward" MS gates can be experimentally convenient as a sign change of $\theta$ generally requires frequency changes of the driving field.
\end{itemize}

\subsection{Circuit decomposition for four-body spin interactions}
Let us now outline the procedure to simulate a coherent time step $e^{\mathcal{L}_\alpha^{\mathrm{coh}} \tau}$ for a four-spin interaction term $H_\alpha = - E_\alpha A_\alpha$ with $A_\alpha = \s{x}_1\s{x}_2\s{x}_3\s{x}_4$ (see figure~\ref{fig:latticemodels}). Although the unitary propagator can in principle be implemented with a standard universal set of single- and two-qubit gates available for ions \cite{T_haffner-physrep-469-155}, here we use an alternative technique, which harvests the multi-ion MS gates and makes use of an ancilla qubit \cite{T_duer-pra-78-052325} encoded in an additional ion (see figure \ref{fig:gatesequence}). This technique has been used in \cite{T_barreiro-nature-2011} to experimentally realize four-body spin interactions.

The approach consists of a sequence of three gate operations: (i) First, a  fully entangling MS gate $U_\mathrm{MS}(\pi/2,0)$, applied to the four system ions and the ancilla ion, coherently maps the information, whether the system ions are in a +1 or -1 eigenstate of $A_x$ onto the ancilla qubit (see figure \ref{fig:gatesequence}(a)). (ii) Second, a single-qubit gate $U_\mathrm{anc}(\phi) = \exp(i \phi \s{z}_0)$ is carried out on the ancilla ion. Due to the previous mapping, this operation on the ancilla qubit is equivalent to manipulations on the +1 and -1 subspaces of $A_\alpha$. (iii) Finally, the mapping is reversed by an inverse MS gate $U_\mathrm{MS}(-\pi/2,0)$ on all ions. The evolution according to the three unitaries is given by
\begin{eqnarray}
\label{eq:three_steps}
U  &= & U_{\rm{MS}}(-\pi/2,0) U_{\rm{anc}}(\phi) U_{\rm{MS}}(\pi/2,0) \nn\\
& = & \exp \left[ i\frac{\pi}{4} \hat{S}_x \s{x}_0 \right]  \exp[i \phi \s{z}_0 ] \exp \left[ -i\frac{\pi}{4} \hat{S}_x \s{x}_0 \right] \nn \\
& = &\exp \left[i \phi \left (\cos\left( \frac{\pi}{2}\hat{S}_x\right) \s{z}_0 + \sin\left( \frac{\pi}{2}\hat{S}_x\right) \s{y}_0  \right) \right]
\end{eqnarray}
with the operator $\hat{S}_x = \sum_{i=1}^n \s{x}_i$ acting on the $n$ system ions. Using the identities
\begin{eqnarray}
\label{eq:cos_identities}
\cos\left( \frac{\pi}{2}\hat{S}_x\right) & = &
\left\{
   \begin{array}{ccl}
     A_\alpha & \textrm{ for } & n = 4k, \, k \in \mathbb{N} \\
    - A_\alpha & \textrm{ for } & n = 4k-2, \, k \in \mathbb{N} \\
     0 & \textrm{ for } & n \textrm{ odd}\\
   \end{array}
\right. \nn \\
\end{eqnarray}
and
\begin{eqnarray}
\label{eq:sin_identities}
\sin\left( \frac{\pi}{2}\hat{S}_x\right) & = &
\left\{
   \begin{array}{ccl}
     A_\alpha & \textrm{ for } & n = 4k-3, \, k \in \mathbb{N} \\
    - A_\alpha & \textrm{ for } & n = 4k-1, \, k \in \mathbb{N} \\
     0 & \textrm{ for } & n \textrm{ even}\\
   \end{array}
\right. \nn \\
\end{eqnarray}
one finds that for $n=4$ eq.~(\ref{eq:three_steps}) indeed reduces to 
\begin{equation}
\label{eq:full_unitary}
U = \exp(i \phi \s{z}_0 \otimes A_\alpha). 
\end{equation}
As a consequence, the ancilla - initially prepared in $\ket{0}$ - factorizes out from the dynamics of the system ions, which in turn evolve according to the unitary time evolution operator $\exp(i \phi A_\alpha)$. Here, from $\exp(i\phi A_\alpha) = \exp(-i(-E_\alpha A_\alpha)\tau)$ one identifies the energy scale of the four-body interaction as $E_\alpha = \phi/\tau$, where $\tau$ is the physical time, which is needed to perform all gates required for one full simulation time step (\ref{eq:Trotter_expansion}). Note that pairwise interactions among the system ions, present in the two-body Hamiltonian underlying the MS gate (\ref{eq:MS_gate}), cancel out in the inverse mapping step (second MS gate).

\subsection{Toolbox for simulation of $n$-body spin interactions}
The simulation scheme outlined above for four-body interactions is readily generalized to arbitrary $n$-body interactions of the form $A = \prod_{i=1}^n \s{\alpha}_i$ with $\s{\alpha}_i \in \{1, \s{x}_i, \s{y}_i, \s{z}_i \}$. This is possible by applying local rotations to (a subset of) system ions before and after the gate sequence and thereby effectively transforming $\s{x}_i$ into $\s{y}_i$ or $\s{z}_i$, and by varying the phase $\phi$ of the MS gates. This enables, e.g., the simulation of $z$-type four-body plaquette interaction terms as required for the toric code Hamiltonian (see figure \ref{fig:latticemodels} and section \ref{sec:kitaev}). The required gate sequences are summarized in table \ref{table:gate_table}. If certain ions are supposed not to participate in the interactions (i.e.~$\s{\alpha}_i =1$) this can be achieved in different ways: (i) by focusing the driving laser of the MS gate operation only onto the relevant subset of ions, or (ii) by hiding the electronic population of these ions in uncoupled electronic states for the duration of the gate sequence \cite{T_roos-science-304-1478}, or (iii) by means of refocusing techniques \cite{T_vandersypen-rmp-76-1037}: As shown in reference \cite{T_nebendahl-pra-79-012312} interspersing MS gates applied to all ions with single-qubit gates on individual ions allows one to decouple effectively certain ions from the dynamics. A set of convenient gate sequences for this purpose is discussed in \ref{app:refocusing}. Circuit decompositions for the simulation of more complex many-spin interactions going beyond $n$-qubit Pauli operators, such as e.g.~ring-exchange interactions, can be worked out and implemented in an analogous way (e.g.~in \cite{T_weimer-nphys-6-382} such cases are discussed). 

We note that in the gate-based ``digital" simulation scheme the energy scale $E_0$ of the $n$-body interactions is essentially independent of the order $n$, and mainly limited by the inverse time required for performing the $(n+1)$-ion MS gates. This is in contrast to analog simulation approaches, where higher-order interactions typically arise in a perturbative limit from a two-body Hamiltonian, thus with correspondingly smaller energy scales. 

\begin{table}[h]
\begin{center}
\begin{tabular}{r|cl}
\multicolumn{2}{c}{} \\
\hline \hline
$A = \prod_{i=1}^n \s{x}_i$ & $ U_{\rm{MS}}(-\pi/2,0) U_{\rm{anc}}(\phi) U_{\rm{MS}}(\pi/2,0)$ \\
\hline 
$n=1, 5, ... $ & $U_\mathrm{anc}(\phi) = \exp(-i\phi \sigma_0^y)$ \\
$n=2, 6, ... $ & $U_\mathrm{anc}(\phi) = \exp(-i\phi \sigma_0^z)$ \\
$n=3, 7, ... $ & $U_\mathrm{anc}(\phi) = \exp(+i\phi \sigma_0^y)$ \\
$n=4, 8, ... $ & $U_\mathrm{anc}(\phi) = \exp(+i\phi \sigma_0^z)$ \\
\hline
$A = \prod_{i=1}^n \s{y}_i$ & $ U_{\rm{MS}}(-\pi/2,\pi/2) U_{\rm{anc}}(\phi) U_{\rm{MS}}(\pi/2,\pi/2)$ \\
\hline 
$n=1, 5, ... $ & $U_\mathrm{anc}(\phi) = \exp(+i\phi \sigma_0^x)$ \\
$n=2, 6, ... $ & $U_\mathrm{anc}(\phi) = \exp(-i\phi \sigma_0^z)$ \\
$n=3, 7, ... $ & $U_\mathrm{anc}(\phi) = \exp(-i\phi \sigma_0^x)$ \\
$n=4, 8, ... $ & $U_\mathrm{anc}(\phi) = \exp(+i\phi \sigma_0^z)$ \\
\hline
\end{tabular}
\caption{\label{table:gate_table} Circuit decompositions for the simulation of one time step of coherent dynamics according to the time evolution operator $U = \exp(i \phi A)$. The unitary block is implemented by two MS gates applied to the $n$ system ions and an ancilla qubit (\# 0) initially prepared in $\ket{0}$, and a single-qubit rotation on the ancilla qubit.}
\end{center}
\end{table}

Finally, we remark that the coherent $n$-body interactions $A_\alpha = \sigma_1^x \ldots \sigma_n^x$ can also be achieved without an ancilla qubit by a slight modification of the employed quantum circuit (see appendix \ref{app:redsequence} for details).

\section{Engineering dissipative many-body dynamics}
\label{sec:diss_simulation}

In this section we show how to engineer dissipative dynamics according to $n$-qubit stabilizer pumping. To be specific, we first discuss the implementation of master equation dynamics governed by four-body plaquette quantum jump operators, $c_\alpha = \frac{1}{2} \sigma_i^z (1- \sigma_1^x \sigma_2^x \sigma_3^x \sigma_4^x)$, as required e.g.~for the ground state cooling of Kitaev's toric code (as discussed above in section \ref{sec:idea_and_results}). The stabilizer pumping, as described in this section, has been  demonstrated in an experiment with five ions \cite{T_barreiro-nature-2011}, four of them representing four system spins, which can be regarded as one plaquette, and one additional ion encoding an ancilla qubit, which has been optically pumped to engineer the dissipative four-spin dynamics.

\subsection{Engineering four-body quantum jump operators for stabilizer pumping}
The dissipative pumping dynamics to ``cool" into the ground state manifold of Kitaev's toric code Hamiltonian, as sketched in figure \ref{fig:plaquette_cooling}, is implemented by three unitary gate operations applied to the four system ions and the ancilla qubit initially prepared in $\ket{0}$, followed by optical pumping of the ancilla qubit. The sequence of unitaries is
\begin{equation}
\label{eq:U_d}
U_d = U_{\rm{MS}}(-\pi/2,0) C(\theta) U_{\rm{MS}}(\pi/2,0)
\end{equation}
with the two-qubit gate 
\begin{equation}
\label{eq:correcting_gate}
C_i(\theta) =\ket{0}\bra{0}_0 \otimes 1 + \ket{1}\bra{1}_0 \otimes \exp(i\theta \s{y}_i).
\end{equation}
(i) As for the coherent simulation, an entangling MS gate $U_\mathrm{MS}(\pi/2,0)$ first maps the information on whether the four system ions are in the +1 or -1 eigenspace of $A_\alpha$ onto the logical states of the ancilla qubit. (ii) Next, the gate $C(\theta)$ realizes a spin flip with a $\theta$-dependent amplitude, provided the ancilla is in $\ket{1}$, i.e.~only if the system spins are in a -1 eigenstate of $A_\alpha$. In \ref{app:gate_decompositions} we give a possible decomposition of $C_i(\theta)$ into global MS gates and single-ion rotations. (iii) After reversing the initial mapping (i) by another (inverse) MS gate, the ancilla qubit is in general entangled with the four system spins. (iv) Finally, optical pumping of the ancilla ion back to its initial state $\ket{0}$ constitutes the dissipative element in the sequence, which renders the dynamics of the four system spins irreversible and enables to carry away entropy and thereby ``cool" the system qubits.

The unitary sequence (\ref{eq:U_d}) can be expressed as $ U_d(\theta) \equiv U_1^\dagger U_0^\dagger C_i(\theta) U_0 U_1$ with $U_0  = \exp (-i (\pi/4) \s{x}_0 \hat{S}_x )$, $\hat{S}_x = \sum_{k=1}^{n=4} \s{x}_k$ and $U_1 = \exp \left(-i (\pi/4) \s{x}_i \sum_{k(\neq i)}^{n=4} \s{x}_k \right)$. 
Here we have made use of the fact that all pairwise interaction terms not involving either the ancilla ion or the $i$-th system ion cancel out due the inverse MS gate. The resulting operation $U_d(\theta) = U_1^\dagger \left[(1+U_3) +  U_2 (1-U_3) \right] U_1 /2$, where $U_2 = \cos ( (\pi/2) \hat{S}_x ) \s{z}_0 + \sin ((\pi/2) \hat{S}_x ) \s{y}_0$ and $U_3 = U_0^\dagger \exp(i\theta \s{y}_i) U_0$,  
can be further simplified using the operator identities (\ref{eq:cos_identities}) and (\ref{eq:sin_identities}) for $n=4$, as well as $U_1^\dagger \s{z}_i U_1 = i \s{z}_i A_\alpha$ and $A_\alpha^2 = 1$, yielding
\begin{equation}
\hspace{-1cm}
U_d(\theta) = \frac{1}{2} (1+ \cos \theta) + \frac{1}{2} \sin \theta \s{z}_i A_\alpha \s{x}_0 + \frac{i}{2} \sin \theta \s{z}_i \s{y}_0 + \frac{1}{2} (1- \cos \theta) A_\alpha \s{z}_0.
\end{equation}
In combination with the subsequent optical pumping of the ancilla ion, the resulting dynamics is given by the quantum operation
\begin{eqnarray}
\label{eq:quantum_operation}
\ket{0}\bra{0}_0 \otimes  \rho_s & \mapsto & \ket{0}\bra{0}_0 \otimes \mathrm{tr}_{0}\{ U_d(\theta) (\ket{0}\bra{0}_0 \otimes \rho_s ) U_d(\theta)^\dagger\} \nn \\
& = & \ket{0}\bra{0}_0 \otimes \sum_{k=1,2} E_k(\theta) \rho_s E_k(\theta)^\dagger
\end{eqnarray}
with the operation elements
\begin{eqnarray}
E_1(\theta)  & = & \frac{1}{2}(1 + A_\alpha) + \cos \theta \, \frac{1}{2} (1 - A_\alpha), \\
E_2(\theta) & = & \sin \theta \, \s{z}_i \frac{1}{2}(1 - A_\alpha) = \sin \theta \, c_\alpha.
\end{eqnarray}
With a probability $p = \sin^2 \theta$ states in the -1 eigenspace of $A_\alpha$ are dissipatively converted into +1 eigenstates, while the +1 eigenspace is left invariant by the operation. For $\theta = \pi/2$ cooling occurs with unit probability.

For small values $\theta \ll \pi/2$ equation (\ref{eq:quantum_operation}) can be expanded up to second order in $\theta$, yielding the standard form of a Lindblad master equation (\ref{eq:Lindblad}) with a four-body jump operator $c_\alpha$ and the corresponding dissipative rate $\gamma_\alpha = \theta^2/\tau$. Here, as above, $\tau$ is the physical time needed for the implementation of one simulation time step (\ref{eq:Trotter_expansion}).

\subsection{Toolbox for dissipative quantum simulation}
The described four-step scheme is readily generalized to $n$-body stabilizer cooling with $n$-qubit quantum jump operators of the form $c = \frac 12 \s{z}_i (1 - A_\alpha)$, where $A_\alpha = \prod_{j=1}^n \s{\alpha}_i$ with $\s{\alpha}_i \in \{1, \s{x}_i, \s{y}_i, \s{z}_i \}$. 
In table \ref{table:dissipative_gates} the required gate operations and the resulting $n$-body quantum jump operators are listed. By combining the outlined scheme with local rotations on (subsets of) the system ions, this allows one, e.g., to engineer cooling dynamics according to $z$-type four-body quantum jump operators $c_\beta = \frac{1}{2} \sigma_i^x (1 - \sigma_1^z \sigma_2^z \sigma_3^z \sigma_4^z)$, which are required for ground state preparation in Kitaev's toric code model, as explained in section \ref{sec:kitaev}.

\begin{table}[h]
\begin{center}
\begin{tabular}{r|llr}
\multicolumn{2}{c}{} \\
\hline \hline
$c = \frac{1}{2} \sigma_i^z (1- \sigma_1^x \ldots \sigma_n^x)$ & $ U_{\rm{MS}}(-\pi/2,0) C_i(\theta) U_{\rm{MS}}(\pi/2,0)$ \\
\hline 
$n=1, 5, ... $ & $ C_i(\theta) = \ket{y_-}\bra{y_-}_0 \otimes 1 + \ket{y_+}\bra{y_+}_0 \otimes \exp(-i\theta \s{z}_i)$ \\
$n=2, 6, ... $ & $ C_i(\theta) = \ket{1}\bra{1}_0 \otimes 1 + \ket{0}\bra{0}_0 \otimes \exp(-i\theta \s{y}_i)$ \\
$n=3, 7, ... $ & $ C_i(\theta) = \ket{y_+}\bra{y_+}_0 \otimes 1 + \ket{y_-}\bra{y_-}_0 \otimes \exp(-i\theta \s{z}_i)$ \\
$n=4, 8, ... $ & $ C_i(\theta) = \ket{0}\bra{0}_0 \otimes 1 + \ket{1}\bra{1}_0 \otimes \exp(i\theta \s{y}_i)$ \\
\hline
\end{tabular}
\caption{\label{table:dissipative_gates} Required gate operations for the simulation of dissipative dynamics according to $n$-body quantum jump operators $c$, which generate pumping into the +1 eigenspace of the many-body stabilizer operator $A = \sigma_1^x \ldots \sigma_n^x$ (and $\sigma_1^y \ldots \sigma_n^y$, respectively). The form of the two-qubit gates $C(\theta)$ are listed for different values of $n$; see \ref{app:gate_decompositions} for convenient decompositions into single-qubit and collective MS gate operations. Here $\ket{y_\pm}$ denote the eigenstates of $\sigma^y$, $\sigma^y \ket{y_\pm} = \pm \ket{y_\pm}$.}
\end{center}
\end{table}

\section{Applications}
\label{sec:applications}
In this section we discuss two examples of possible realizations of complex spin models within the presented simulation scheme. We start with a few comments on boundary effects related to the simulation of minimal instances of Kitaev's toric code Hamiltonian \cite{T_Kitaev-annalsphys-303-2}. Subsequently, we proceed with the discussion of a minimal instance of a recently suggested color code model \cite{T_bombin-prl-97-180501}, whose implementation is feasible in a setup of eight ions. The simulation of other models, typically involving similar many-body spin interaction terms, poses comparable demands with regard to the simulation abilities of coherent and dissipative dynamics. 

\subsection{Kitaev's toric code}
\label{sec:kitaev}

\begin{figure}[ht]
\begin{center}
\includegraphics[width=0.8\columnwidth]{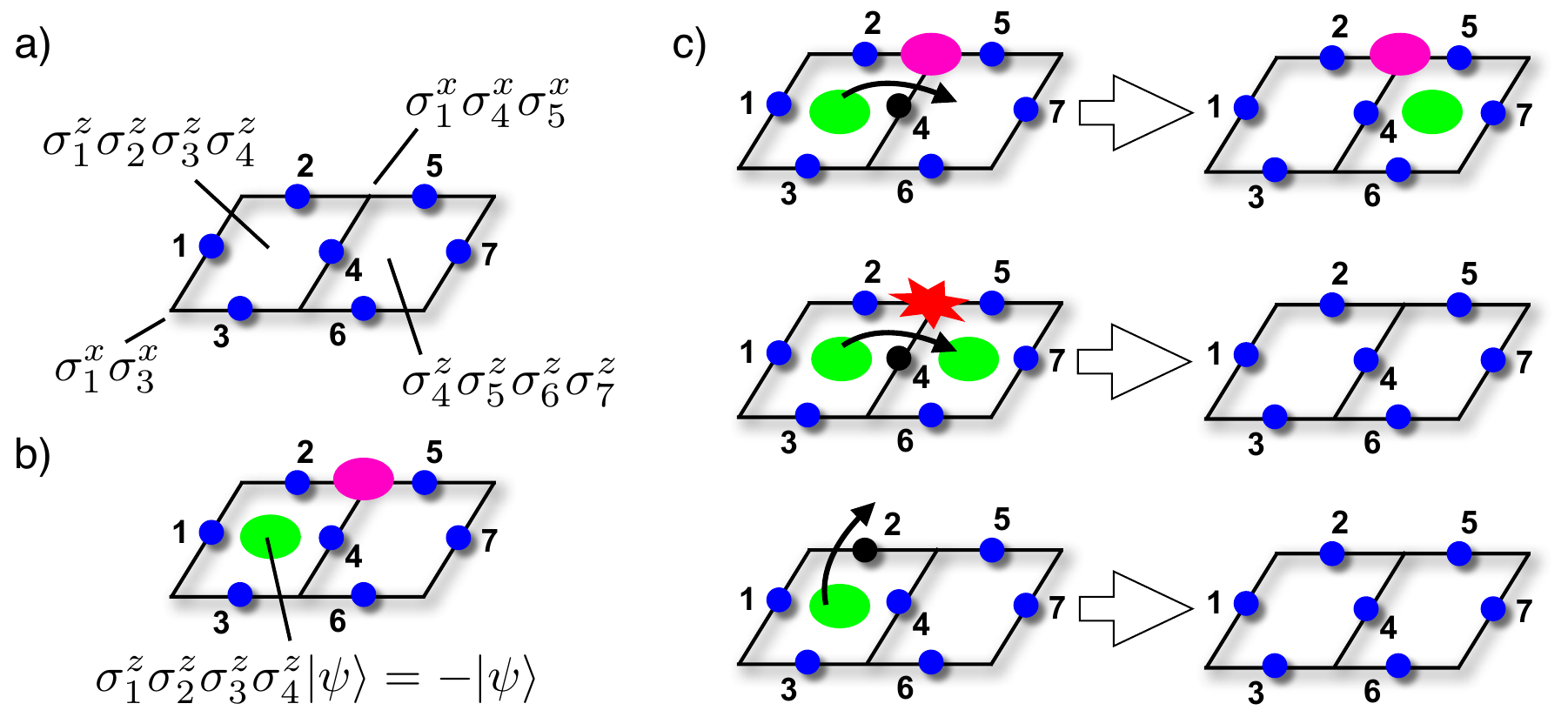} 
\end{center}
\caption{Cooling dynamics in a small-scale implementation of Kitaev's toric code. (a) Two- and three-qubit stabilizer operators for the boundary spins. (b) Two types of excitations in the model correspond to violations (eigenstates of eigenvalue -1) of $x$ and $z$-type stabilizer operators. (c) Cooling dynamics due to pairwise ``annihilation" of excitations or single excitations which are pushed out of the system.}
\label{fig:2KitaevPlaquettes} 
\end{figure}

For small instances of the toric code model, as e.g.~a system of two plaquettes as illustrated in figure \ref{fig:2KitaevPlaquettes}(a), it is possible to define reduced two- or three-body stabilizers for the spins located at the border of the system. For this small system of two plaquettes, coherent and dissipative dynamics can be implemented in a setup of of 7+1 ions.
Dissipative dynamics that cools into the ground state manifold of the model, is realized by a Liouvillian with four-body quantum-jump operators as given in eq.~(\ref{eq:Kitaev_operators}). As discussed above, they realize pumping into the low-energy subspaces, i.e.~eigenspaces corresponding to an eigenvalue +1 of the plaquette and vertex terms of the Hamiltonian.
In figure \ref{fig:2KitaevPlaquettes}(c) the effect of a quantum jump induced by the operator $c = \frac 12 \sigma_4^x (1 -\sigma_1^z \sigma_2^z \sigma_3^z \sigma_4^z)$ is illustrated. If an excitation on the left plaquette is present, in a quantum jump the state of the system spins is of the converted into state, which is a +1 eigenstate of the left plaquette operator. In this process the excitation hops over to the neighboring plaquette. If a second excitation was present on the second plaquette, the pair of excitations is ``annihilated" in such step. This removal of a pair of excitation constitutes a cooling event. Note that the other type of excitation remains unaffected by this dynamics. Alternatively, cooling takes also place if one of the spins which forms the border of the system is flipped (e.g.~spin \# 2 in the lower part of figure \ref{fig:2KitaevPlaquettes}(c)), since then a single excitation is ``pushed out" from the lattice system. 
For the corners and edges of a small lattice system two- and three-body jump operators for $x$-type stabilizer pumping can be realized in analogy.

\subsection{Simulation of a color code model}
\label{sec:colorcode}

The idea of storing and processing quantum information in naturally protected quantum systems has attracted a lot of interest in recent years \cite{T_Kitaev-annalsphys-303-2,T_nayak-revmodphys-80-1083}. Here, protection from local errors is achieved by encoding quantum information not in individual physical qubits, but instead in ground states of topologically ordered quantum systems, which provide an energy gap to excited states and exhibit a ground state degeneracy, which cannot be lifted local perturbations. One class of topological quantum error correcting codes that have been proposed in this context, are color codes \cite{T_bombin-prl-97-180501}, which exhibit remarkable computational and error correcting capabilities. In particular, they allow us to implement the Clifford group in a fully topological way within the ground state manifold, without the need of individual addressing of physical qubits or braiding of quasiparticles.

Here, we outline how a minimal instance of such a color code can be realized with the discussed simulation tools, and discuss the state preparation and implementation gate operations, as well as readout. For a detailed introduction to color code models we refer the reader to \cite{T_bombin-pra-76-012305}.

A minimal version of a color code, consisting of seven physical qubit located at the corners of three plaquettes, is shown in figure \ref{fig:latticemodels}(b). Including one ancilla qubit for the implementation of coherent and dissipative dynamics this system could be simulated with a \textit{string of eight ions}. Qubits located around plaquettes interact via four-body $x$ and $z$-interaction terms: The Hamiltonian $H_\mathrm{cc} = - E \left( \sum_{i=1}^3 A_i + \sum_{i=1}^3 B_i\right)$ with $A_1 = \s{x}_1\s{x}_2\s{x}_3\s{x}_4$, $B_1 = \s{z}_1\s{z}_2\s{z}_3\s{z}_4$, and similar interaction terms for the other two plaquettes, consists of six mutually commuting stabilizer operators. 

Coherent dynamics according to this Hamiltonian can be realized by implementing unitary time steps as outlined in section \ref{sec:coh_simulation} with the help of an ancilla qubit. Cooling into the ground state manifold can be achieved by a Liouvillian dynamics associated with a set of six four-body jump operators, such as $c = \frac 12 \s{z}_1(1- \s{x}_1\s{x}_2\s{x}_3\s{x}_4)$, thereby driving the system spins dissipatively into the +1 eigenspaces of the six stabilizers. Alternatively, ground state cooling can be realized by a sequence of deterministic cooling steps: Starting with the system spins in the fully polarized state $\ket{0}^{\otimes 7}$ (being already a +1 eigenstate of all $z$-stabilizers) it suffices to implement three dissipative Kraus maps (\ref{eq:Kraus_map}) such as 
\begin{equation}
\rho \mapsto \frac{1}{2} (1+ A) \rho \frac{1}{2} (1+ A) + \frac{1}{2} \s{z}_1 (1- A) \rho \frac{1}{2} (1- A) \s{z}_1
\end{equation}
with $A = \s{x}_1\s{x}_2\s{x}_3\s{x}_4$
and accordingly for the other two plaquettes, in order to prepare the system also in the +1 subspaces of the $x$-type stabilizer operators. As discussed in section \ref{sec:diss_simulation} this is achieved by choosing two-qubit correcting gates $C_i(\theta = \pi/2)$ (see equation~(\ref{eq:correcting_gate})). 

Excited states $\ket{\psi}$ of the Hamiltonian $H_\mathrm{cc}$ correspond to states, where the system spins are in -1 eigenstate(s) with respect to certain stabilizer(s). The quasiparticles of $x$ (or $z$) type associated with these violations of the stabilizer constraints are located on the corresponding plaquettes, for instance on the uppermost plaquette, if $A_1 \ket{\psi} = - \ket{\psi}$ (or $B_1 \ket{\psi} = - \ket{\psi}$).

Since there are only six stabilizer constraints for seven system spins, the ground state is degenerate and thus offers the possibility to encode one logical qubit. 
An operator basis for this \textit{logical} qubit can be constructed by the global operators $\hat{X} = \prod_{i=1}^7 \s{x}_i$ and $\hat{Z} = \prod_{i=1}^7 \s{z}_i$.
These two logical operators commute with all six stabilizers of the code, thus they leave the system within the ground state manifold.

The logical qubit can be initialized in the logical state $\ket{\bar{0}}$ by (dissipatively) preparing the system - in analogy with the four-body stabilizer cooling - in a +1 eigenstate of the global operator $\hat{Z}$, such that $\hat{Z} \ket{\bar{0}} =\ket{\bar{0}}$. The logical $\ket{\bar{1}}$-state is then obtained by the application of the logical $\hat{X}$-operator, $\ket{\bar{1}} =\hat{X} \ket{\bar{0}}$, which corresponds to a single-qubit rotation applied to all seven system ions. This minimal color code setup also allows to implement single-qubit gates belonging to the Clifford group in a topological way: The Hadamard $H$ and phase-shift gate $K$
\begin{equation}
H = \frac{1}{\sqrt{2}} \left( \begin{array}{cc} 1 & 1 \\ 1 & -1 \end{array} \right), \qquad K = \left( \begin{array}{cc} 1 & 0 \\ 0 & i \end{array} \right),
\end{equation}
can be implemented by applying the corresponding operations globally to all seven system ions, i.e.~$\hat{H} = \prod_{i=1}^7 H_i$ and $\hat{K} = \prod_{i=1}^7 K_i$. The logical operators then directly fulfill the required transformation properties, as for example $\hat{H}^\dagger \hat{X}\hat{H} = \hat{Z}$.

Remarkably, once the system is prepared in the code space, the logical single-qubit gates can be performed by simple global single-qubit rotations without the need of addressing individual ions. These operators take ground states to ground states, the system stays in the code space, and thus the quantum gates are achieved without braiding of quasiparticles. Similarly, readout measurements can be performed globally, i.e.~by standard fluorescence imaging of all ions measured either in the $x$ or in the $z$-basis.

For the realization of a topological C-NOT gate operation, the minimal system consists of two seven-qubit-layers encoding two logical qubits. Its implementation therefore requires an experimental setup of fifteen ions, and might thus become experimentally feasible in the near future.

\section{Noise and Imperfections}
\label{sec:imperfections}

In a stroboscopic simulation of a many-body master-equation (\ref{eq:master_equation}) several sources of imperfections occur. First of all, Trotter errors from the non-commutativity of coherent (and also dissipative) terms arise for each time step of the simulation. These are bounded and can be reduced by resorting to smaller time steps and higher-order Trotter decompositions. In addition, imperfect gate operations in the quantum circuits lead to errors. Their effects on gate-based quantum simulations has discussed in detail e.g.~in reference \cite{T_duer-pra-78-052325}. 

\subsection{Generic effect of gate imperfections on the quantum simulation}
Here, we first briefly discuss the generic effect by considering a particularly transparent example of a pulse length error in the simulation of coherent dynamics $U = \exp(i \phi A)$ according to a four-body spin interaction $A = \sigma_1^x \sigma_2^x \sigma_3^x \sigma_4^x$, as explained in section \ref{sec:coh_simulation}. We assume that the only error is a pulse length error in the single-qubit gate $U_\mathrm{anc}(\phi) = \exp(i\phi \sigma_0^z)$ applied to the ancilla qubit in the three-step sequence eq.~(\ref{eq:three_steps}). In one small time step ($\phi \ll \pi/2$), the system spins evolve according to
\begin{equation}
\rho( t + \tau) \simeq U \rho(t) U^\dagger \simeq \rho(t) - i [-\phi A, \rho(t)] + \phi^2 (A \rho(t) A - \rho(t)) 
\end{equation}
Assuming that the actual value $\phi$ fluctuates (e.g.~due to laser intensity fluctuations) in the experiment from time step to time step according to a Gaussian distribution $p(\phi) = 1/\sqrt{2 \pi \sigma^2} \exp[-(\phi-\phi_0)^2/(2 \sigma^2)]$ around the mean value $\phi_0$ with a variance $\sigma \ll \phi_0$ we obtain, after averaging over $\phi$, the modified equation of motion
\begin{equation}
\label{eq:errors}
\frac{\mathrm{d}}{\mathrm{d}t} \rho \simeq - i [(-\phi_0/\tau) A, \rho] + \frac{\phi_0^2}{\tau} (A \rho A - \rho) + \frac{\sigma^2}{\tau}(A \rho A - \rho)
\end{equation}
Thus, one finds dynamics according to a four-body Hamiltonian $H_\mathrm{eff} = - (\phi_0/\tau) A$, where a systematic shift in $\phi$ results in a systematically larger or smaller energy scale. In addition, the stochastic Gaussian fluctuations in $\phi$ cause a collective dephasing dynamics (in the $\sigma^x$-basis), described by a Liouvillian with a dephasing rate $\gamma = \sigma^2/\tau$ and a hermitian four-body quantum jump operator $A$ (see last term in (\ref{eq:errors})).The effect of other gate errors in the circuit decompositions for coherent and dissipative dynamics can be analyzed in an analogous way. A more specific error analysis, going beyond these quite general arguments, requires more precise information about the dominant error sources in concrete experimental setups.

\subsection{Comparison with experimental stabilizer pumping}
\begin{figure}[h!]
\begin{center}
\vspace{3mm}
\includegraphics[width=0.95\columnwidth]{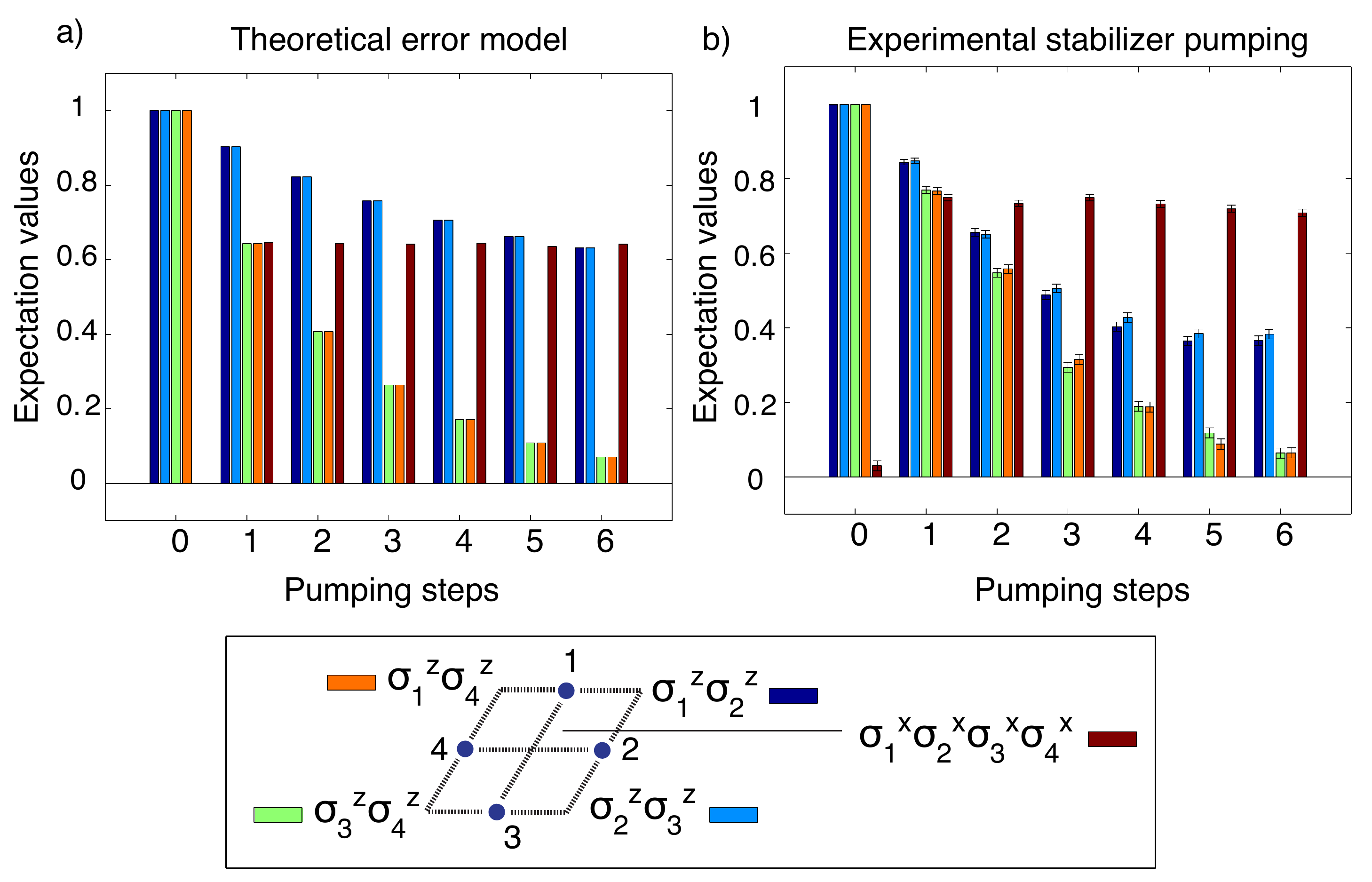}
\end{center}
\caption{\textbf{a)} Results from a numerical simulation, which accounts for the effect of single-qubit gate errors in repeated stabilizer pumping according to the dissipative map (\ref{eq:Kraus_map_repeated}). The stabilizer expectation values are obtained by averaging over 10000 realizations, assuming un-correlated errors from gate to gate, in the phases of the single-qubit rotations $\exp(-i(\theta + \delta \theta)/2 \sigma_{0,4}^z)$ with $\delta \theta$ obeying a Gaussian distribution with zero mean and a variance of $\delta \theta$ of $0.3 \times \pi/2$. \textbf{b)} Experimental repeated stabilizer pumping. Plot reproduced from data given in the supplementary information of the Nature publication \cite{T_barreiro-nature-2011}. Quantitative differences from the numerical simulation are mainly due to additional errors in the global gate operations, whose precise form is unknown and which have not been taken into account in the theoretical error model.}
\label{fig:repeated_pumping}
\end{figure}

In the work \cite{T_barreiro-nature-2011} four-qubit stabilizer pumping and the effect of errors have been studied experimentally. For the benefit of the reader and to make the present discussion
self-contained we find it worthwhile to review briefly the main findings, as
explained in detail in the supplementary information of the Nature article \cite{T_barreiro-nature-2011}, to relate this to the present discussion. In the experiment with five ions (which encoded four system qubits and one additional ancilla qubit) stabilizer pumping with 100\% pumping probability per step, from the -1 into the +1 eigenspace of the four-qubit stabilizer operator $A = \sigma_1^x \sigma_2^x \sigma_3^x \sigma_4^x$ has been applied repetitively. The corresponding discrete Kraus map reads $\rho \mapsto E_1 \rho E_1^\dagger + E_2 \rho E_2^\dagger$ with operation elements
\begin{equation}
\label{eq:Kraus_map_repeated}
E_1 = \frac{1}{2} (1+ \sigma_1^x \sigma_2^x \sigma_3^x \sigma_4^x) \qquad \mathrm{and} \qquad E_2 = \frac{1}{2} \s{z}_4 (1- \sigma_1^x \sigma_2^x \sigma_3^x \sigma_4^x).
\end{equation}
Starting with the four system qubits in the initial state $\ket{1111}$, ideally these reach the four-qubit GHZ state $(\ket{0000}+\ket{1111})/\sqrt{2}$ after a single application of the above Kraus map. This is reflected by the fact that the expectation value of the four-qubit stabilizer $A$ assumes a value of +1 after the application of this dissipative step. At the same time, the expectation values of the two-qubit stabilizer operators $\sigma_i^z \sigma_j^z$, as depicted in schematically in figure \ref{fig:repeated_pumping}, should ideally remain unaffected by the four-qubit stabilizer pumping dynamics and stay at a value +1. 

In the experiment, the Kraus map (\ref{eq:Kraus_map_repeated}) has been realized by a quantum circuit consisting of \textit{global} rotations and MS gate operations applied to all five ions and \textit{addressed} $z$-type single-qubit rotations, which only involve the ancilla qubit (index \# 0) and the system qubit with index \# 4 (see the supplementary information of \cite{T_barreiro-nature-2011} for the exact form of the experimental circuit decomposition). It is reasonable to assume that errors in the global gate operations affect all ions to a similar degree, whereas gate errors in the addressed single-qubit gates lead to additional errors, which dominantly act on the system qubit \# 4. As a consequence, one expects that under a repeated application of the dissipative map (\ref{eq:Kraus_map_repeated}), the expectation values of the two-qubit stabilizer operators, which involve $\sigma_4^z$ to decay faster than those not involving the ion \#4. This decay behavior, as found from a numerical simulation (see figure \ref{fig:repeated_pumping}(a) and figure caption for details), has qualitatively been observed in the experiment \cite{T_barreiro-nature-2011}. In addition to the errors in the single-qubit gates (which experimentally are rooted in pulse length errors and / or laser intensity fluctuations, as discussed in the previous subsection), the dominant source of errors are imperfections in the MS gate operations, which result in quantitative differences in the observed decay of stabilizer expecation values. In the language of stabilizer models, these errors in the simulation correspond to unwanted heating processes with respect to the $z$-type stabilizers during the four-body stabilizer pumping according to the map (\ref{eq:Kraus_map_repeated}).

\section{Conclusions and Outlook}
\label{sec:summary_and_outlook}

In this work we have discussed a toolbox for ``digital" quantum simulation with linear chains of trapped ions. We have outlined the theoretical concepts and details of the experiment, which recently demonstrated the building blocks of an open-system quantum simulator with up to five ions. Furthermore, we have discussed how our scheme allows one to explore the physics and simulate the coherent and dissipative dynamics of minimal instances of spin models involving $n$-body interactions and constraints, such as e.g.Kitaev's toric code and a minimal version of topological color code model. Similarly, circuit implementations for more complex coherent and dissipative $n$-body interaction terms as, e.g., plaquette exchange interactions can be developed; see for instance \cite{T_weimer-nphys-6-382}.

\begin{figure}[ht]
\begin{center}
\includegraphics[width=0.7\columnwidth]{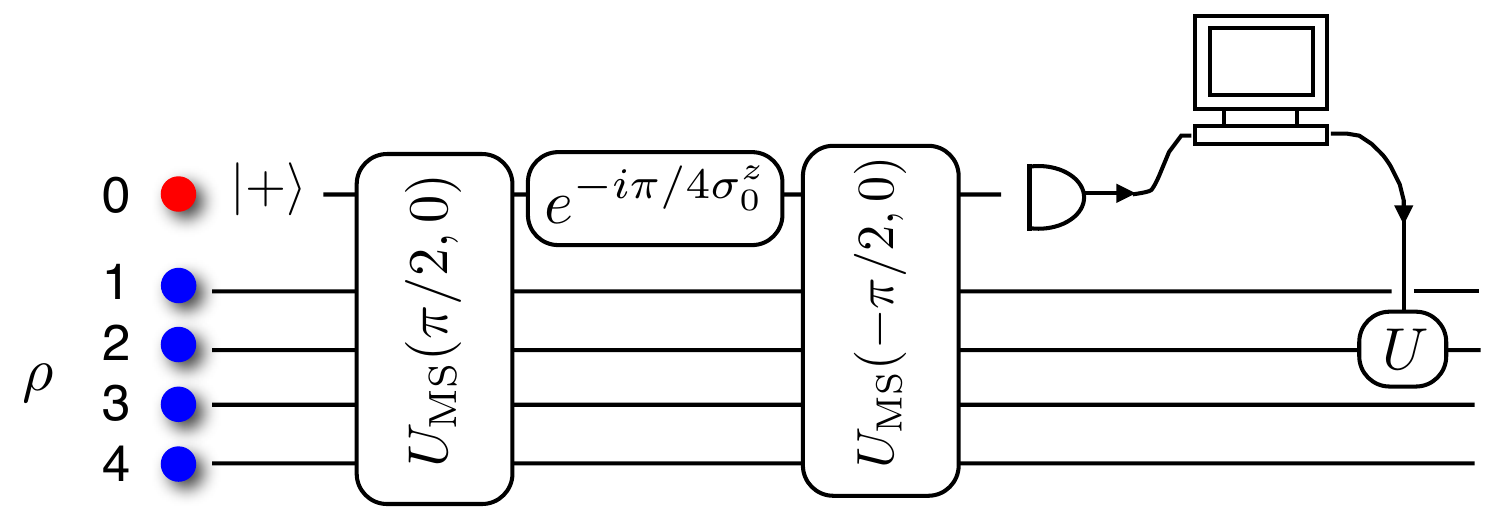}
\end{center}
\caption{Circuit for readout of the four-body stabilizer operator $A = \sigma_1^x\sigma_2^x\sigma_3^x\sigma_4^x$ via a measurement of the ancilla qubit. The circuit for coherent simulation of a four-body spin interactions as discussed in section \ref{sec:coh_simulation} (cf.~eq.~(\ref{eq:full_unitary})) realizes the unitary operation $U = \exp(iÊ\pi/4 \sigma_0^z \otimes A)$. Thereby, the ancilla qubit initially prepared in $\ket{+}$ is coherently mapped onto the two $\sigma^y$-eigenstates, depending on whether the four system qubits are in a +1 or -1 eigenstate of $A$. This information obtained from a subsequent measurement of the ancilla qubit in the appropriate basis can be classically processed and used for feedback operations on the system.
}
\label{fig:ancilla_measurement}
\end{figure}

Here, we have focused on \textit{open-loop} dynamics, where coherent and dissipative time evolution in stabilizer models is implemented with the aid of an ancilla qubit, which is not observed. It is known that such open-loop dynamics involving a single, non-observed ancilla qubit is not sufficient to realize the most general Markovian multi-qubit open-system dynamics. As shown in \cite{T_lloyd-pra-65-010101}, this can be achieved by a \textit{closed-loop} simulation scenario. Here, general open-system quantum operations are realized by consecutive sequences of coherent operations applied to the system qubits and the ancilla, interspersed with measurements of the ancilla qubit in an appropriate basis. The gathered information from the outcomes of the sequential ancilla measurements can be classically processed and used for feedback operations on the system. We note that the described scheme also allows one to extract such information about the system qubits via a measurement of the ancilla qubit, as schematically shown in figure \ref{fig:ancilla_measurement}. 
In addition, the measurement of $n$-body observables such as multi-qubit stabilizer operators is an essential ingredient e.g.~for error syndrome measurements in quantum error correction and quantum computing protocols \cite{T_steane-prl-77-793,T_calderbank-pra-54-1098,T_dennis-j-mat-phys-43-4452}. 

The engineering of reservoir couplings and dissipative many-body processes enables novel directions for quantum state preparation \cite{T_diehl-natphys-4-878}, as recently also shown in an experiment with atomic ensembles \cite{T_krauter-arxiv-1006.4344}. Combining dissipative time evolution with coherent Hamiltonian dynamics might allow one to explore novel physics such as non-equilibrium phase transitions in driven dissipative systems \cite{T_diehl-prl-105-015702}. In particular, the ability to implement e.g.~master equations with multi-qubit quantum jump operators opens interesting perspectives for building quantum memories based on dissipation \cite{T_pastawski-arxiv:1010.2901} or the demonstration of a novel form of quantum computing solely based on dissipation \cite{T_verstraete-nphys-5-633}.

\ack
We would like to thank H Bombin, M A Martin-Delgado and W D\"ur for discussions. We also acknowledge discussions with the experimental ion-trap group of R Blatt at Innsbruck. We acknowledge support by the Austrian Science Fund (FOQUS), the European Commission (AQUTE) and the Institute of Quantum Information. Y L Zhou acknowledges support by the Hunan Provincial Innovation Foundation for Postgraduate and NSFC Grant No.~11074307.

\appendix

\section{Coherent Dynamics without an Ancilla Qubit}
\label{app:redsequence}

\begin{figure}[ht]
\begin{center}
\includegraphics[width=0.95\columnwidth]{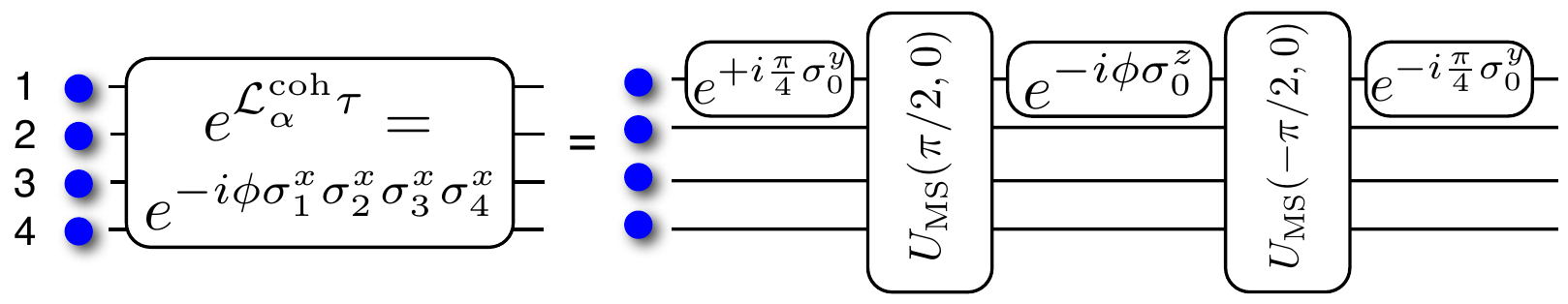}
\end{center}
\caption{Quantum circuit for a coherent time step according to a four-spin Hamiltonian term   $A_\alpha = \sigma_1^x \ldots \sigma_n^x$, without using an extra ancilla qubit.}
\label{fig:red_sequence}
\end{figure}

Coherent dynamics according to $n$-body spin interactions of the form $A_\alpha = \sigma_1^x \ldots \sigma_n^x$ can also be achieved without an ancilla qubit as follows: 
By inspection of equation (\ref{eq:full_unitary}) one sees that the quantum circuit involving the $n$ system qubits (e.g. $n=4$) and the single ancilla qubit actually realizes coherent time evolution $\exp(i\phi \sigma_0^z \otimes A_\alpha)$, according to an $(n+1)$-body spin interaction term $\sigma_0^z \sigma_1^x \ldots \sigma_n^x$. This evolution is -- up to a single-qubit rotation of the ancilla qubit around the $y$-axis -- equivalent to evolution according the $(n+1)$-body interaction term $\sigma_0^x \sigma_1^x \ldots \sigma_n^x$. In other words, for coherent $n$-body interactions it suffices that one of the $n$ system qubits takes the role played by the ancilla. The resulting quantum circuit is shown in figure \ref{fig:red_sequence}.
%

\section{Refocusing Techniques}
\label{app:refocusing}
As outlined in section \ref{sec:coh_simulation}, MS gates on subsets of ions can be achieved (i) either by transferring ions, which are not supposed to participate in the gate, into decoupled electronic levels, by the application of hiding pulses, or (ii) alternatively by employing refocusing techniques. In this appendix we review how MS gates on subsets of ions can be achieved by decomposing the desired unitary operations into sequences of MS gates, which are applied to \textit{all} ions, combined with single-ion refocusing pulses on individiual ions (see also \cite{T_nebendahl-pra-79-012312}).

\textbf{Sequence for a MS gate on $\mathbf{n-1}$ out of $\mathbf{n}$ ions:} 

\begin{figure}[ht]
\begin{center}
\includegraphics[width=0.75\columnwidth]{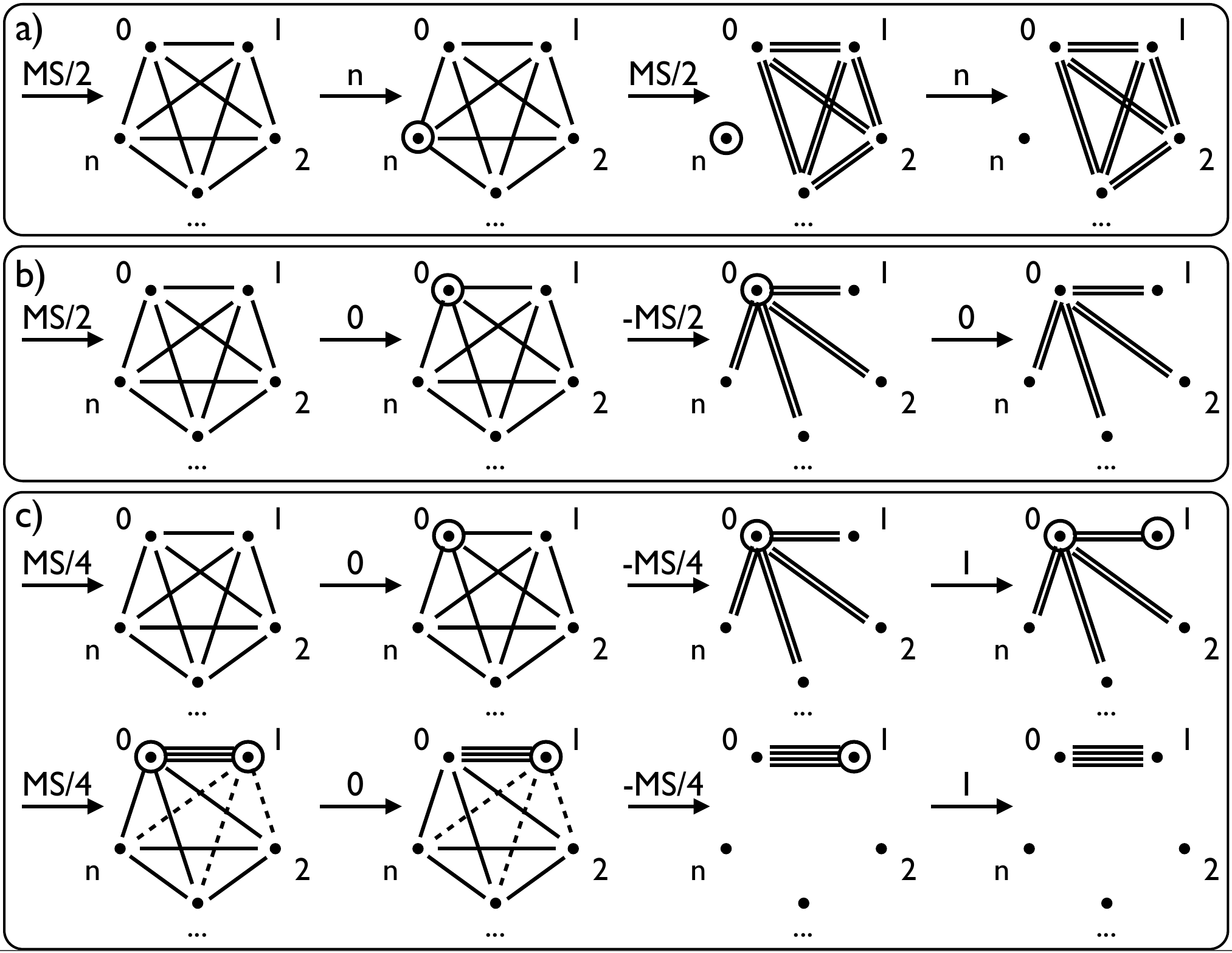} 
\end{center}
\caption{Gate sequences for the realization of entangling gates on subset of ions, by a combination of MS gates applied to all ions and refocusing pulses on individual ions. The nodes represent the $n+1$ ions, lines between nodes $i$ and $j$ denote entanglement created between a pair $\{i,j\}$ of ions during the application of MS gates. The short-hand notation $\pm$MS/2 and $\pm$MS/4 stands for $U_\mathrm{MS}(\pm\theta/2,\phi)$ and $U_\mathrm{MS}(\pm\theta/4,\phi)$, respectively. An operation $j$ denotes a single ion pulse $U_{\s{z}}^{(j)}(\pi)$ applied to the $i$-th ion; ions which have been exposed to such flip operations are labelled by a small circle, until they are flipped back into their original orientation. \textbf{a)} Gate decomposition for a MS on all ions except the $n$-th ion (cf.~eq.~(\ref{eq:MS_n_minus_1})). \textbf{b)} Gate sequence for the creation of star-type entanglement between the auxiliary ion (\#0) and each of the $n$ system ions. \textbf{c)} Gate sequence for a MS gate on two ions out of $n+1$. Dashed lines correspond to entanglement, which is created in intermediate steps between an initially disentangled pair of ions $\{i,j\}$, if one (and only one) of the two ions (marked with a circle) has been previously flipped.}
\label{fig:refocusing} 
\end{figure}

A MS gate $U_\mathrm{MS}(\theta,\phi)$ on all but, say, the $n$-th ion can be implemented by a combination of two MS gates of half of the angle $\theta$, and two single-ion $z$-gates $U_{\s{z}}^{(n)} = \exp(-i \pi/2 \s{z}_n)$ applied to the $n$-th ion, i.e.
\begin{equation}
\label{eq:MS_n_minus_1}
U_\mathrm{MS}^{(0,1,...,n-1)}(\theta,0)  = U_{\s{z}}^{(n)}(\pi) \, U_\mathrm{MS}(\theta/2,\phi)\,U_{\s{z}}^{(n)}(\pi) \,U_\mathrm{MS}(\theta/2,\phi)
\end{equation}
up to an irrelevant global phase. The sequence of four gates can be understood as follows: With the first MS gate ``half" of the final entanglement is created between all pairs $\{i,j\}$ of the $n$ ions, due to the pairwise interaction terms underlying the MS gate (\ref{eq:MS_gate}). Now, the spin of the $n$-th ion is flipped by $U_{\s{z}}^{(n)}(\pi)$, such that in what follows $\s{x}_n$ and $\s{y}_n$ act effectively as $-\s{x}_n$ and $-\s{y}_n$. In the third step, the second ``half" MS gate then entangles all pairs of ions, which do not include the $n$-th ion, further; only for the pairs of ions which involve the $n$-th ion the entanglement creation of the first step is reversed. In this way, the $n$-th ion is effectively decoupled from all other $n-1$ ions. Finally the $n$-th is flipped back into its initial orientation by another single-qubit gate $U_{\s{z}}^{(n)}(\pi)$. The four steps are graphically illustrated in figure \ref{fig:refocusing}a. 

\textbf{Star-type MS gate of the auxiliary ion with all system ions:}
It is also straightforward to realize an entangling gate between the ancilla ion and each system ion without creating pairwise entanglement between the system ions. This can be done by the sequence
\begin{equation}
\prod_{i=1}^n U^{(0,i)}_\mathrm{MS}(\theta,\phi) = U_{\s{z}}^{(0)}(\pi) \, U_\mathrm{MS}(-\theta/2,\phi)\,U_{\s{z}}^{(0)}(\pi) \,U_\mathrm{MS}(\theta/2,\phi)
\end{equation}
which is sketched in figure \ref{fig:refocusing}b. Here, the second \textit{inverse} MS gate, which is applied after the single qubit flip of the auxiliary ion, cancels the initially generated entanglement between all pairs of ions, which do not include the ancilla ion.

\textbf{Sequence for a MS gate on $\mathbf{2}$ of $\mathbf{n}$ ions:}
The sequence for the implementation of the star-type entangling operation discussed in the previous paragraph can be used to realize a MS gate on two of $n$ ions (see figure \ref{fig:refocusing}c). Such two-ion MS gate is an essential building block for the implementation of the two-qubit gate $C(\theta)$, which is needed for the dissipative simulation discussed in section \ref{sec:diss_simulation}. For instance, the sequence for a MS gate on only the auxiliary  ion and the system ion \#1 is given by
\begin{eqnarray}
\label{eq:MS_on_2}
U_\mathrm{MS}^{(0,1)}(\theta, \phi)  & = & U_{\s{z}}^{(1)}(\pi) \, U_\mathrm{MS}(-\theta/4,\phi)\,U_{\s{z}}^{(0)}(\pi) \,U_\mathrm{MS}(\theta/4,\phi) \nn \\
&& \times U_{\s{z}}^{(1)}(\pi) \, U_\mathrm{MS}(-\theta/4,\phi)\,U_{\s{z}}^{(0)}(\pi) \,U_\mathrm{MS}(\theta/4,\phi)
\end{eqnarray}

More involved decompositions for MS gates, where more than one or two ions are supposed to participate in or be excluded from the gate operation, can be constructed accordingly. In general, they will involve more ``partial" MS gates and refocusing pulses, which might at some point render the alternative approach of hiding pulses on individual ions more suitable.

\section{Gate Decompositions}
\label{app:gate_decompositions}

Here, we provide decompositions of the two-qubit gates $C_i(\theta)$, which are needed for the dissipative $n$-body dynamics as discussed in section \ref{sec:diss_simulation}, into MS gates and single-ion rotations.

For the simulation of $n$-body interactions with $n=4, 8, ...$ the gate $C_i(\theta)$ of Eq.~(\ref{eq:correcting_gate}) can be decomposed as
\begin{eqnarray} 
C_i(\theta) & = &  \ket{0}\bra{0}_0 \otimes 1 + \ket{1}\bra{1}_0 \otimes \exp[i\theta \s{y}_i] \nn \\
& = & e^{\frac 12 (1- \s{z}_0) i \theta \s{y}_i} \nn \\
& = & e^{\frac{i \theta}{2}\s{y}_i}  \, e^{-\frac{i \theta}{2} \s{z}_0 \s{y}_i} \nn \\
& = &  e^{\frac{i \theta}{2}\s{y}_i}  \, e^{-\frac{i \pi}{4}\s{x}_0} \, U_\mathrm{MS}^{(0,i)}(\theta/2, \pi/2) \,e^{\frac{i \pi}{4}\s{x}_0}
\end{eqnarray}
The two-qubit MS gate on the auxiliary ion and the $i$-th system ion $U_\mathrm{MS}^{(0,i)}(\theta/2, \pi/2)$ can be realized via refocusing techniques - see eq.~(\ref{eq:MS_on_2}) in \ref{app:refocusing}.

It is straightforward to decompose the two-qubit gates $C_i(\theta)$ for other values of $n$ (as listed in table \ref{table:dissipative_gates}) accordingly.

\vspace{5mm}
\bibliography{manuscript.bbl}


\end{document}